\documentclass[fleqn,usenatbib]{mnras}

\usepackage[T1]{fontenc}
\usepackage{ae,aecompl}

\usepackage{times}

\usepackage{graphicx}	
\usepackage{amsmath}	
\usepackage{amssymb}	

\newcommand{\vect}[1]{\boldsymbol{#1}}
\newcommand{\diff}{\text{d}}
\newcommand{\vel}[1]{\mathrm{#1}}

\date{}
\pubyear{2017}
\title[Tides in superfluid neutron star binaries]{Resonant tidal excitation of superfluid neutron stars in coalescing binaries}

\author[H. Yu and N. N. Weinberg]{Hang Yu$^{1, 2}$\thanks{E-mail: hyu45@mit.edu} and Nevin N. Weinberg$^{1}$\\
$^{1}$Department of Physics, and Kavli Institute for Astrophysics and Space Research, Massachusetts Institute of Technology, \\Cambridge, MA 02139, USA\\
$^{2}$LIGO Laboratory, Massachusetts Institute of Technology, Cambridge, MA 02139, USA}

\begin{document}

\label{firstpage}
\pagerange{\pageref{firstpage}--\pageref{lastpage}}
\maketitle

\begin{abstract}
We study the resonant tidal excitation of g modes in coalescing superfluid neutron star (NS) binaries and investigate how such tidal driving impacts the  gravitational-wave (GW) signal of the inspiral.  Previous studies of this type treated the NS core as a normal fluid and thus did not account for its expected superfluidity. The source of buoyancy that supports the g modes is fundamentally different in the two cases: in a normal fluid core, the buoyancy is due to gradients in the proton-to-neutron fraction, whereas in a superfluid core it is due to gradients in the muon-to-electron fraction.  The latter yields a stronger stratification and a superfluid NS therefore has a denser spectrum of g modes with frequencies above $10\textrm{ Hz}$.  As a result, many more g modes undergo resonant tidal excitation as the binary sweeps through the bandwidth of GW detectors such as LIGO.  
We find that $\simeq 10$ times more orbital energy is transferred into g mode oscillations if the NS has a superfluid core rather than a normal fluid core. However, because this energy is transferred later in the inspiral when the orbital decay is faster,  the accumulated phase error in the gravitational waveform is comparable for a superfluid and a normal fluid NS ($\sim 10^{-3}-10^{-2}\textrm{rad}$). A phase error of this magnitude is too small to be measured from a single event with the current generation of GW detectors. 
\end{abstract}

\begin{keywords}
binaries: close -- stars: interiors -- stars: neutron -- stars: oscillations.
\end{keywords}

\section{INTRODUCTION}

Advanced LIGO's detection of the merger of  binary black holes  heralds a new age of gravitational-wave (GW) astronomy \citep{Abbott:16, Abbott:16b}.  Coalescing binary neutron star (NS) systems and NS-black hole systems, although not yet detected \citep{Abbott:16c}, are another promising source for ground based GW detectors such as Advanced LIGO, Advanced Virgo, and KAGRA (respectively, \citealt{Harry:10, Acernese:15, Somiya:12}).  The rich array of  science that their detection might deliver (for a recent review see \citealt{Baiotti:16}) includes the exciting prospect of constraining the enigmatic supranuclear equation of state from measurements of the tide-induced phase shift of the GW signal \citep{Read:09, Hinderer:10, Damour:12, Lackey:12, Lackey:15, Agathos:15}. 

The linear tidal response of the NS can be decomposed into an equilibrium tide and a dynamical tide. The equilibrium tide accounts for the quasi-static, large scale distortion of the star and the dynamical tide accounts for the internal modes of oscillation that are resonantly excited as the orbit decays and sweeps up in frequency.  While most recent studies focus on the impact of the equilibrium tide on the GW signal (including all the references listed at the end of the previous paragraph), there is also an extensive literature studying the impact of the linear dynamical tide.  \citet{Lai:94} and \citet{Reisenegger:94} considered non-rotating normal fluid NSs, where the resonant modes are g modes with frequencies $\lesssim 100\textrm{ Hz}$. They found that the excited g modes only weakly affect the GW signal (phase shifts of $\lesssim 10^{-2}\textrm{ radian}$; see also \citealt{Shibata:94, Kokkotas:95}).  Subsequent studies accounted for rotation and found that a rapidly rotating NS  could have a much stronger tidal response, resulting in phase shifts of $\sim 0.1$ to $\gg 1$ radian \citep{Ho:99, Lai:06, Flanagan:07}. However, this requires a spin frequency $\gtrsim \textrm{ a few} \times 100\textrm { Hz}$, which is larger than is thought to be likely for a NS in a coalescing binary \citep{Brown:12}. Most recently, \citet{Hinderer:16} developed an effective-one-body waveform model that accounts for the resonant response of the high frequency f-modes. They found that in some cases the f-mode contribution to the phase shift might be as much as $\approx 30\%$ of the total tidal effect.  
  
All of these studies assumed a normal fluid NS.  However,  because the NSs in coalescing binaries are expected to be cold, the core neutrons will be a superfluid  (\citealt{Yakovlev:99, Lombardo:01}). The source of buoyancy that provides the restoring force for g modes is fundamentally different for normal fluid and superfluid NSs. In a normal fluid NS, a perturbed fluid element is buoyant due to gradients in the proton-to-neutron fraction  \citep{Reisenegger:92}. However, in a superfluid NS the neutrons within the fluid element can flow past the protons and gradients in their relative abundance no longer provides buoyancy.  Indeed, studies that assume a zero temperature superfluid NS composed only of neutrons, protons, and electrons (and not muons) find that such stars do not support  g mode oscillations (e.g., \citealt{Lee:95, Andersson:01, Prix:02}).\footnote{The focus here is on g modes supported by composition gradients.  At finite temperatures, thermal gradients are also a source of buoyancy  \citep{Gusakov:13}.  However, for the cold NSs in coalescing binaries,  thermal gradients make a negligible contribution to the total buoyancy \citep{Passamonti:16}.}

More recently, \citet{Kantor:14} showed that when the presence of muons is taken into account,  there is a new source of buoyancy in the core: the gradient in the muon-to-electron fraction.  Thus, a cold superfluid NS does support core g modes when we extend the model to include a richer chemical composition.

Since the source of buoyancy is different, the g modes of a superfluid NS are different from the g modes of a normal fluid NS.  In particular, \citet{Kantor:14} showed that the stratification is  considerably stronger in a superfluid NS, i.e., the Brunt-V{\"a}is{\"a}l{\"a} frequency is larger (see also \citealt{Passamonti:16}). As a result, the entire g mode spectrum is shifted to higher frequencies, including the $l=2$ g modes that are resonantly excited by the tide.  We will show that a superfluid NS has more than ten $l=2$ g modes with frequency $>50 \textrm{ Hz}$  whereas a normal fluid NS has only two or three such modes. This means that there are many more g modes that undergo resonant excitation as the binary sweeps through the bandwidth of ground-based detectors such as LIGO.  Moreover, the nature of the tidal coupling is different in a superfluid NS since the tide forces not one but two fluids (the neutron superfluid and the normal fluid consisting of the charged particles).  The purpose of our study is to account for these superfluid effects and thereby extend previous calculations of the  dynamical tide in NS binaries.

 The plan of the paper is as follows: in Section \ref{sec:bgNS} we describe our background superfluid NS model and discuss the source of buoyancy in more detail.  In Section \ref{sec:Formalism} we describe tidal driving in superfluid NSs beginning with a calculation of the stellar eigenmodes.  In Section  \ref{sec:energyPhase} we present the main result of our study, the calculation of the GW phase shift induced by the resonant excitation of g modes. In Section \ref{sec:conclusion} we summarize and conclude.

\section{SUPERFLUID NEUTRON STAR MODEL}\label{sec:bgNS}

We construct our background superfluid NS models using an approach that is similar to that of \cite{Prix:02} except that we account for the existence of muons in the core.  This is an important distinction since, as already mentioned in the introduction and described further in Section \ref{sec:buoyancy}, the muon-to-electron composition gradient provides the buoyancy that supports g modes in the core. 

We assume an NS composed of neutrons ($\textrm{n}$), protons ($\textrm{p}$), electrons ($\textrm{e}$) and muons ($\mu$), and adopt the SLy4 equation of state for baryons \citep{Stone:03}, while treating the leptons as relativistic degenerate Fermi gas.  Since an NS in a coalescing binary is expected to be  cold ($T\ll 10^8 \textrm{ K}$), we neglect thermal effects (we set $T=0$) and assume that the neutrons in the core are superfluid. In the crust, taken to be the region with baryon density $n_\text{b} < 0.1 \textrm{fm}^{-3}$, we treat all species of particles as normal fluid matter for simplicity, which is consistent with the treatment in \citeauthor{Kantor:14} (2014; see also \citealt{Dommes:16}).  In order to simplify the calculation of the oscillation modes (Section \ref{sec:Formalism}), we neglect rotation and use Newtonian equations throughout our analysis including, for consistency, in constructing the background hydrostatic models. Corrections to the stellar and mode structure due to general relativistic effects are expected to be at the level of $GM/(Rc^2) \sim 20\%$, where $M$ and $R$ are the mass and radius of the NS. Such corrections are unlikely to change the overall conclusions of our study.  We assume all charge densities are strictly balanced and  neglect all electrodynamic effects (including proton superconductivity, plasma oscillations, and magnetic fields). We also neglect vortex-tension and vortex pinning of superfluid neutrons, as is appropriate for the macroscopic description of fluid flow that is of interest here.  For a more detailed discussion of these effects and the underlying assumptions, we refer the reader to \cite{Prix:02} and references therein. 

Given the above simplifications, we can describe the NS as consisting of two fluids: a normal fluid of charged particles (protons, electrons, and muons) and a superfluid of neutrons whose flow drifts through the normal fluid flow.  We indicate the fluid variables of the charged (neutron) flow with a subscript $\rm{c}$ ($\rm{n}$).   The dynamics of the flow depends on the total internal energy density of the cold superfluid, which is given by \citep{Prix:02}
\begin{equation}
\diff \varepsilon_{\textrm{tot}} = \sum_{j=\textrm{npe}\mu} \mu_j \diff n_j+ \alpha \diff \vel{v}_{\rm r}^2,
\end{equation}
where $n_j$ and $\mu_j$ are particle $j$'s number density and chemical potential, respectively, with $j$ being one of n, p, e, or $\mu$. The quantity $\vect{\vel{v}}_\textrm{r}$ is the relative velocity between the normal fluid (charged) flow and the superfluid (neutron) flow, 
\begin{equation}
\vect{\vel{v}}_{\rm r} = \vect{\vel{v}}_{\rm c} - \vect{\vel{v}}_{\rm n}
\end{equation}
and $\alpha$ is the entrainment function (see below). The pressure of the fluid is given by
\begin{align}
\diff P & = \sum_{j=\textrm{npe}\mu}  n_j \diff \mu_j- \alpha \diff \vel{v}_\textrm{r}^2 \nonumber \\
&=\sum_{j=\textrm{npe}\mu}  \rho_j \diff \tilde{\mu}_j- \alpha \diff \vel{v}_\textrm{r}^2 \nonumber \\
&=\rho_\textrm{c} \diff \tilde{\mu}_\textrm{c} + \rho_\textrm{n} \diff \tilde{\mu}_\textrm{n}  -\alpha \diff \vel{v}_\textrm{r}^2,
\label{eq:dP}
\end{align}
where, for each particle species $j$, we define the mass density $\rho_j = \varepsilon_j/c^2$, the energy density $\varepsilon_j$ (rest mass plus interaction/kinetic energy), the specific chemical potential $\diff \tilde{\mu}_j= \diff \mu_j/m_j$, and the (relativistic) mass $m_j = \rho_j/n_j$.
In the third line we combined the protons, electrons, and muons together to represent our charged flow, with
\begin{align}
&\rho_\textrm{c}=\rho_\textrm{p} + \rho_\textrm{e} + \rho_\mu, \\
&\rho_\textrm{c} \diff \tilde{\mu}_\textrm{c} = \sum_{j=\textrm{pe}\mu} \rho_j \diff \tilde{\mu}_j.
\end{align} 
In Appendix \ref{subsec:basicQuan} we discuss these quantities in more detail and provide some additional thermodynamic relations that we use in our study. 

The $\alpha \diff \vel{v}_{\rm r}^2$ term characterizes the entrainment effect which, in the zero-temperature limit, is due entirely to the strong interaction between neutrons and protons. The entrainment function $\alpha$ can be written as  \citep{Prix:02}
\begin{equation} 
2\alpha = \rho_\textrm{c}\left[1-\frac{m_\textrm{p}^\ast}{m_\textrm{N}} + O\left(\frac{\rho_\textrm{c}}{\rho}\right)\right],
\label{eq:alpha}
\end{equation}
where $m_\textrm{p}^{\ast}$ is the proton effective mass and $\rho=\rho_\textrm{n}+\rho_\textrm{c}$ is the total mass density. Typical values of $m_\textrm{p}^\ast $ are in the range $0.3 \le m_\textrm{p}^{\ast}/m_{\rm N} \le 0.8$ \citep{Sjoberg:76, Chamel:08}. While in general $m_\textrm{p}^{\ast}$ depends on density, for simplicity we consider models that have constant $m_\textrm{p}^{\ast}$ throughout the star. As we will see later, tidal coupling depends only weakly on entrainment effects. 
It will also be useful to describe the entrainment in terms of the dimensionless entrainment functions
\begin{equation}
\epsilon_\textrm{c}=\frac{2\alpha}{\rho_{\rm c}}, \hspace{0.5cm}
\epsilon_\textrm{n}=\frac{2\alpha}{\rho_{\rm n}}.
\end{equation}
We discuss the entrainment function in more detail in Appendix \ref{subsec:Entrainment}.

\begin{figure}
\hspace*{-0.3cm}  
\includegraphics[angle=0,scale=.41]{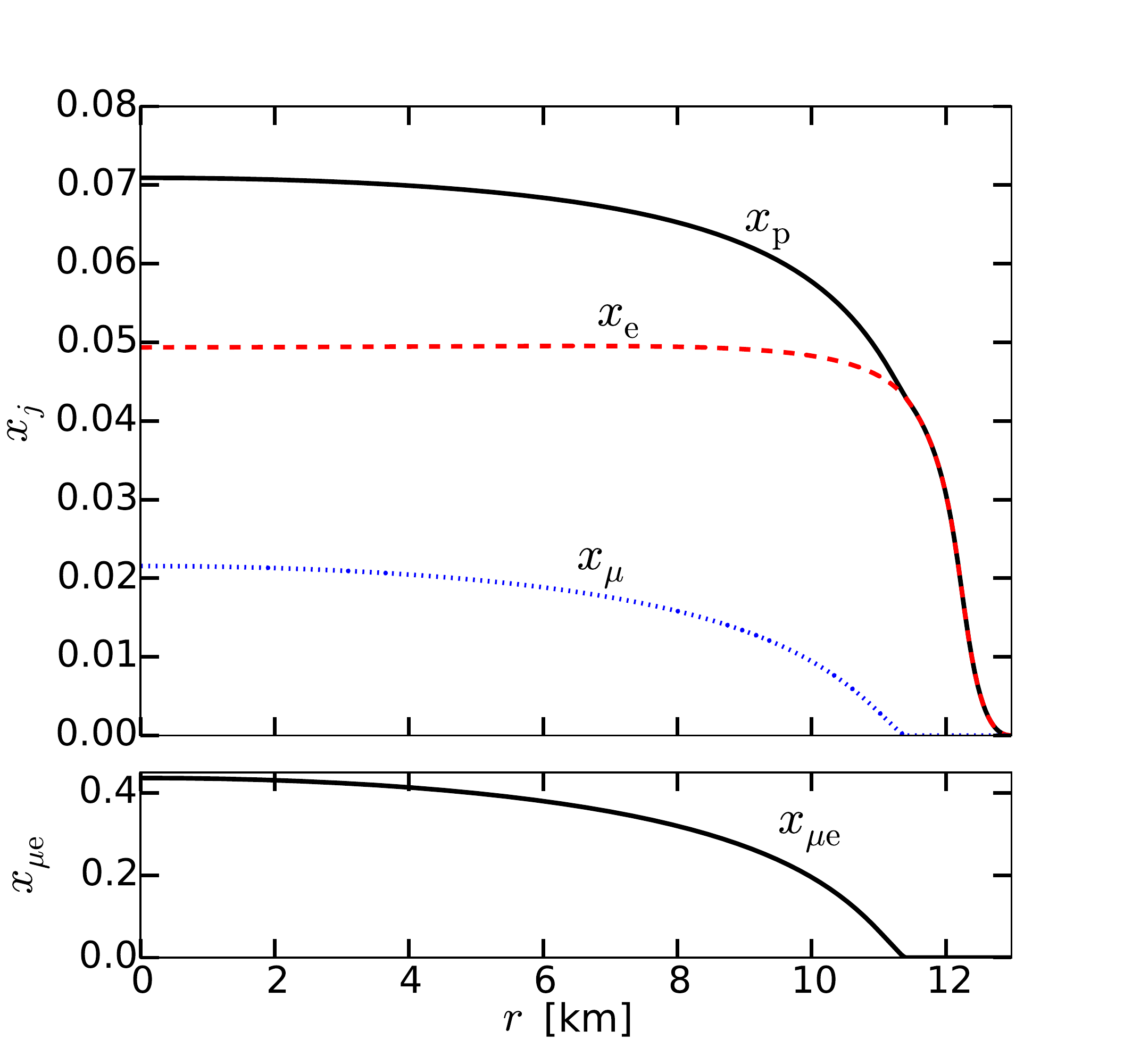}
\caption{Number fraction of protons $x_\textrm{p}$ (solid line),  electrons $x_\textrm{e}$ (dashed line), and muons $x_\mu$ (dotted line), as a function of radius $r$ for our 1.4 $M_\odot$ NS model. In the bottom panel we show the muon-to-electron ratio $x_{\mu{\rm e}}=x_\mu/x_{\rm e}$, which determines the buoyancy profile $\mathcal{N}(r)$ of the superfluid NS model.}
\label{fig:compGrad}
\end{figure}

Using the above relations, we construct spherically symmetric background models by simultaneously solving the equation of hydrostatic equilibrium 
\begin{equation}
\frac{\diff P}{\diff r}=-\left(\rho_{\rm n}+\rho_{\rm c}\right)\frac{\diff \Phi}{\diff r}
\label{eq:hse}
\end{equation}
and chemical (beta) equilibrium
\begin{align}
&\mu_{\rm n}  =\mu_{\rm p} + \mu_{\rm e},\label{eq:baryonBetaEq}\\ 
&\mu_{\rm e} = \mu_\mu,\label{eq:leptonBetaEq}
\end{align}
where $\Phi(r)$ is the gravitational potential and $\mu_e=\mu_\mu$ applies at radii where $\mu_\textrm{e} > m_\mu c^2$ (corresponding to $r<R_\mu$, where $R_\mu$ is the critical radius where muons first appear). Equations (\ref{eq:dP}) and (\ref{eq:hse})-(\ref{eq:leptonBetaEq}) imply 
\begin{equation}
\frac{\diff \tilde{\mu}_\textrm{n}}{\diff r}=-\frac{\diff \Phi}{\diff r}.
\end{equation}
We consider models with masses of 1.4 $M_\odot$ and 2.0 $M_\odot$ and various levels of entrainment. In Table \ref{tab:bgConfig} we give the following parameters of the hydrostatic structure: total mass $M$, radius $R$, central density $\rho_0$, the radius below which muons are present $R_\mu$, and the radius of the core-crust interface $R_{\rm cc}$.  Note that the radii and central density differ from the values in \citet{Stone:03} because we solve the Newtonian hydrostatic equations instead of the general-relativistic equations. In Fig. \ref{fig:compGrad} we show the radial profile of the number fraction $x_j(r)=n_j/(n_\textrm{p} + n_\textrm{n})$ of protons, electrons, and muons and the muon-to-electron ratio $x_{\mu \rm e}(r)=x_\mu/x_{\rm e}$ for the $1.4 M_\odot$ model (the profile of the $2.0 M_\odot$ model is very similar). We observe that the composition varies slowly with radius over most of the star but quickly drops to zero when the radius is close to the critical radius (note that at $R_\mu$ the muon number density goes to zero with a non-zero derivative). 

\begin{table}
\begin{center}
\caption{\label{tab:bgConfig}Parameters of the background NS models.}
\begin{tabular}{ccccc}
\hline
$M$ [$M_\odot$] & $R$ [km]& $\rho_0$ [$10^{14}$ g$\ $ cm$^{-3}$] & $R_\mu$ [km] & $R_{\rm cc}$ [km] \\
\hline
1.4 & 13.0 & 6.7 & 11.4  &11.7 \\
2.0 & 13.7 & 7.85 & 12.5 & 12.7 \\
\hline
\end{tabular}
\end{center}
\end{table}

\subsection{Buoyancy in cold neutron stars}
\label{sec:buoyancy}
Because we assume a zero-temperature NS, composition gradients are the only possible source of buoyancy (i.e., the Ledoux convective stability criterion).  First consider a normal fluid NS consisting of  npe matter. In this case, the buoyancy force that supports the g modes is due to the proton-to-neutron composition gradient (\citealt{Reisenegger:92}; see also \citealt{Lai:94}). To verify this, consider a fluid element in equilibrium with pressure $P$, proton number fraction $x_{\rm p}$ ($=x_{\rm e}$ by charge neutrality), and density $\rho(P,\ x_\textrm{p})$. If we adiabatically displace the element upwards against gravity by a distance $\diff r$, it will remain in near pressure equilibrium with the surroundings by contracting or expanding on a dynamical timescale (which is much shorter than the buoyancy oscillation timescale).  However, its composition will still be $x_\textrm{p}$ because the timescale to reach chemical equilibrium through weak interactions  (which are responsible for changes to $x_\textrm{p}$) is much longer than the buoyancy timescale and because all species of particles within the element move at the same speed \citep{Reisenegger:92}.  The convective stability criterion is therefore
\begin{equation}
\left(\frac{\partial \rho}{\partial x_\textrm{p}}\right)_P\left(\frac{\diff x_\textrm{p}}{\diff r}\right) < 0, 
\end{equation}
where the subscript $P$ indicates the derivative is taken at constant pressure.

\begin{figure*}
\includegraphics[angle=0,scale=.42]{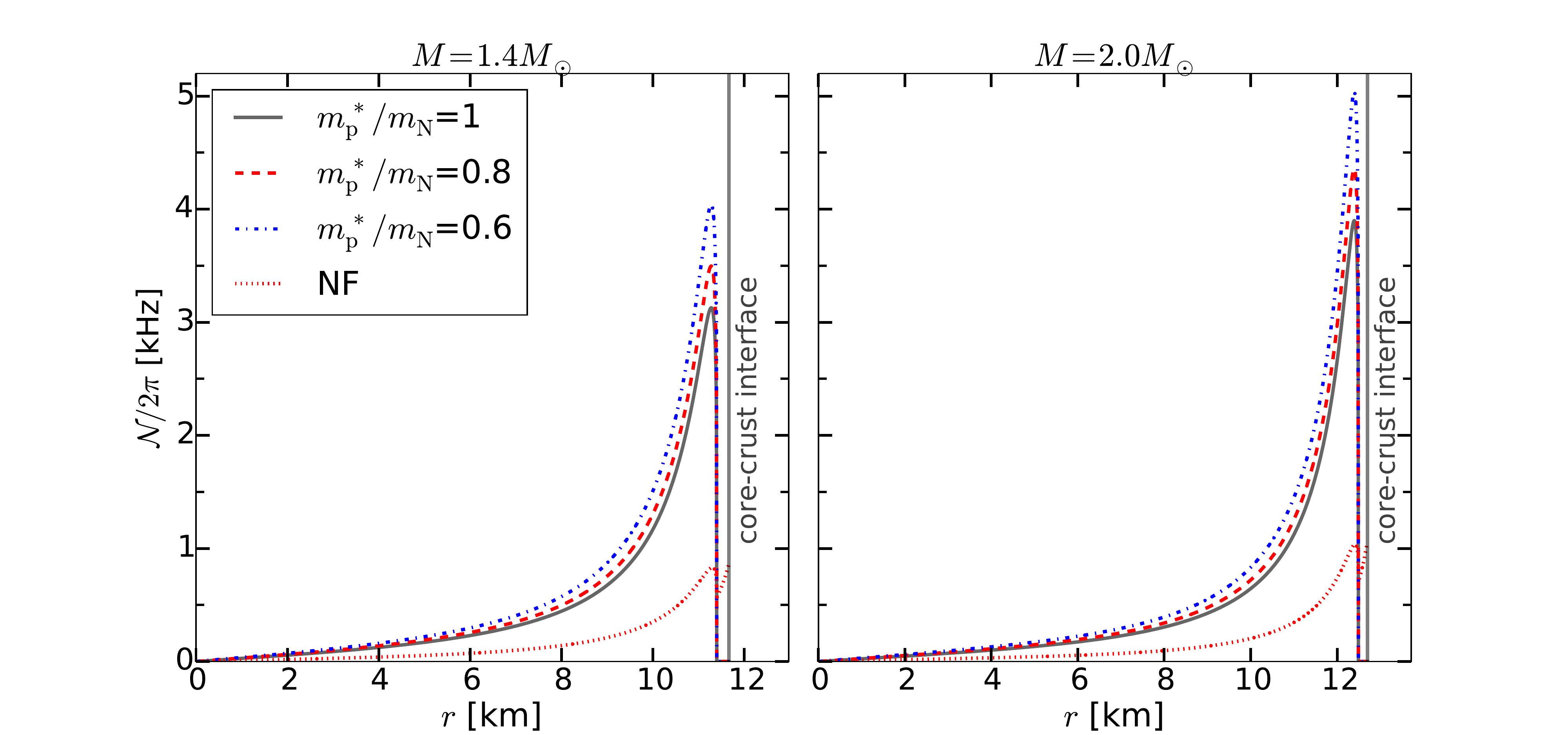}
\caption{Buoyancy frequency $\mathcal{N}/2\pi$ as a function of radius for our $1.4 M_\odot$ (left panel) and $2.0 M_\odot$ (right panel) NS models. We show results for three different entrainment levels, labelled according to their proton effective mass: $m_\textrm{p}^\ast/m_\textrm{N}=1,\ 0.8,\ 0.6$ (black solid lines, red dashed lines, and blue dash-dotted lines, respectively).  We also show $\mathcal{N}/2\pi$ for the normal fluid models (red dotted lines). The vertical lines indicate the core-crust interface. }
\label{fig:BVsimple}
\end{figure*}

Now consider a superfluid NS consisting of only npe matter (no muons). The above stability criterion is no longer valid because the superfluid neutrons form a separate component that is free to drift through the charged components when the fluid element is displaced. This allows the fraction of superfluid neutrons within the element to always match the background (i.e., $x_\textrm{p}$ is not fixed).  As a result, there is no longer a source of buoyancy to support g mode oscillations, as a number of studies have shown (see, e.g., \citealt{Lee:95, Andersson:01, Prix:02}). 

However, the situation changes again when we consider a superfluid NS consisting of npe$\mu$ matter.  There are now three independent variables that parametrize the equation of state.  As in \cite{Kantor:14}, we take these to be $P$, $\mu_\textrm{n}$, and $x_{\mu \textrm{e}} = x_\mu/x_\textrm{e}$. Now if we displace our fluid element, $P$ and $\mu_\textrm{n}$ adjust themselves to the new background values (by contracting/expanding and by varying the number of superfluid neutrons, respectively).  However,  $x_{\mu\textrm{e}}$ remains fixed because the electrons and muons move with the same velocity, that of the charged flow $\vect{\vel{v}}_\textrm{c}$. The stability criterion is therefore
\begin{equation}
\left(\frac{\partial \rho}{\partial x_{\mu\textrm{e}}}\right)_{P,  \mu_\textrm{n}} \left(\frac{\diff x_{\mu\textrm{e}}}{\diff r}\right) < 0, 
\end{equation}
i.e., gradients in $x_{\mu\textrm{e}}$ provide a buoyancy force that can support g modes.

The convective stability criteria given above are closely related to the Brunt-V\"{a}is\"al\"a buoyancy frequency $\mathcal{N}$.  In a npe$\mu$ normal fluid, the density can be uniquely parameterized in terms of $P$, $x_\textrm{e}$, and $x_\mu$ and the buoyancy is given by  
\begin{equation}
\label{eq:N2_normal fluid}
\mathcal{N}^2=-\frac{g}{\rho}\sum_{j=\textrm{e}, \mu}\left[\frac{\partial \rho(P,\ x_\textrm{e},\ x_\mu)}{\partial x_j}\right]_{P,\;x_{i \ne j}} \left(\frac{\diff x_j}{\diff r}\right).
\end{equation}
where $g=\diff\Phi/\diff r$ is the gravitational acceleration.
In a npe$\mu$ superfluid, the density can be uniquely parameterized in terms of $P$, $\mu_\textrm{n}$, and $x_{\mu\rm e}$ and the buoyancy is given by
\begin{equation}
\label{eq:N2_superfluid}
\mathcal{N}^2=-\frac{1-\epsilon_\textrm{n}}{x_\textrm{p} - \epsilon_\textrm{n}}\frac{g}{\rho}\left[\frac{\partial \rho(P,\ \mu_\textrm{n},\ x_{\mu\textrm{e}})}{\partial x_{\mu\textrm{e}}}\right]_{P,\mu_\textrm{n}} \left(\frac{\diff x_{\mu\textrm{e}}}{\diff r}\right)
\end{equation}
 (see, e.g.,  \citealt{Passamonti:16} equations 67, 132, B29, and B38; in Appendix \ref{subsec:Entrainment} we describe how to relate our notation to that used in  \citealt{Passamonti:16}). In Fig. \ref{fig:BVsimple} we show the buoyancy profiles $\mathcal{N}(r)$ of our superfluid and normal fluid models.  The curves are for different combinations of NS mass and entrainment levels; specifically, we show results for a superfluid NS with ($M/M_\odot$, $m_\textrm{p}^\ast/m_\textrm{N}$)=(1.4, 1), (1.4, 0.8), (1.4, 0.6), (2.0, 0.8) and for a normal fluid NS with $M=1.4 M_\odot$. 

We find that $\mathcal{N}(r)$ is a factor of approximately $x_\textrm{p}^{-1}\approx 4$ larger in the superfluid models compared to the normal fluid models (with a mild dependence on stellar mass). This is consistent with the results of \cite{Kantor:14} and \cite{Passamonti:16} (see their Figs. 2 and 6, respectively). Physically, this is because the neutron component is nearly decoupled from the charged component and thus the mass of the oscillating fluid element is smaller by a factor of $\simeq x_\textrm{p}$ compared to the normal fluid case (see equations (\ref{eq:N2_normal fluid}) and (\ref{eq:N2_superfluid}); note that the differential terms in these two equations happen to be comparable). From equations (\ref{eq:alpha}) and (\ref{eq:N2_superfluid}) we also see that a smaller $m_\textrm{p}^\ast$ (that is, a larger $\epsilon_\textrm{n}$), yields a larger $\mathcal{N}$. We will see in Section \ref{sec:inhomog} that the larger $\mathcal{N}$ of the superfluid models shifts the $g$-mode spectrum to higher frequencies.

Note that for $r>R_\mu$ , there are no muons and $\mathcal{N}=0$ in the superfluid case. Finally, for simplicity we neglect the buoyancy of the crust and set $\mathcal{N}=0$ for $r>R_{\rm cc}$ (since only a small fraction of the NS mass is in the crust, this simplification should not significantly affect the core g modes of interest here). 
 
\section{Tidal driving}\label{sec:Formalism}

We now consider small amplitude perturbations to the static background described in Section \ref{sec:bgNS}.  In Section \ref{sec:inhomog} we describe the homogeneous linear eigenvalue problem in which the perturbations are free to oscillate at their natural frequency (i.e., they are not driven by an external force).  In  Section \ref{sec:homog} we describe the inhomogeneous tidal problem in which the perturbations are linearly forced by the tidal potential of the NS's  companion. 

\subsection{Eigenmodes of a superfluid neutron star}
\label{sec:inhomog}

The linearized Newtonian fluid equations describing the free oscillations of the superfluid neutrons and the charged normal fluid are \citep{Prix:02}
\begin{align}
&\vect{\nabla}\cdot (\rho_\textrm{c} \vect{\xi}_\textrm{c}) + \delta \rho_\textrm{c}=0, \label{eq:mass1}\\
&\vect{\nabla}\cdot(\rho_\textrm{n} \vect{\xi}_\textrm{n}) + \delta \rho_\textrm{n}=0, \label{eq:mass2}\\
&\sigma^2\left[\vect{\xi}_\textrm{c} - \epsilon_\textrm{c}(\vect{\xi}_\textrm{c}-\vect{\xi}_\textrm{n}) \right] = \vect{\nabla}\left(\delta \tilde{\mu}_\textrm{c} + \delta \Phi\right), \label{eq:mom1}\\
&\sigma^2\left[\vect{\xi}_\textrm{n} + \epsilon_\textrm{n}(\vect{\xi}_\textrm{c}-\vect{\xi}_\textrm{n}) \right] = \vect{\nabla}\left(\delta \tilde{\mu}_\textrm{n} + \delta \Phi\right), \label{eq:mom2}\\
&\nabla^2\delta \Phi=\delta \rho_\textrm{c} + \delta \rho_\textrm{n}, \label{eq:poisson}
\end{align}
where we assume that the perturbed quantities have a time dependence $e^{i\sigma t}$, $\delta \mathcal{Q}(\vect{x})$ denotes the Eulerian perturbation  of a quantity $\mathcal{Q}$ at location $\vect{x}$, and $\vect{\xi}_\textrm{c}(\vect{x})$ and $\vect{\xi}_\textrm{n}(\vect{x})$  are the Lagrangian displacement fields of the charged normal fluid and neutron superfluid. These equations express mass continuity (eqs. \ref{eq:mass1} and \ref{eq:mass2}), momentum conservation (eqs. \ref{eq:mom1} and \ref{eq:mom2}), and Poisson's equation (\ref{eq:poisson}) relating the perturbed gravitational potential $\delta \Phi$ to the perturbed total density (we do not make the Cowling approximation).

We solve these equations using standard techniques of stellar oscillation theory.  In particular, we consider spheroidal modes in which the perturbed solutions separate into radial and angular functions 
\begin{align}
&\delta \mathcal{Q}(r,\theta,\phi)=\delta \mathcal{Q}(r) Y_{lm}(\theta,\phi), \label{eq:SHdecompScalar}\\
&\vect{\xi}_\textrm{c}(r,\ \theta,\ \phi)=\left[\xi_\textrm{c}^r(r),\ \xi_\textrm{c}^h(r)\frac{\partial}{\partial \theta},\ \xi_\textrm{c}^h(r)\frac{1}{\sin \theta}\frac{\partial}{\partial \phi}\right] Y_{lm}(\theta,\ \phi)
\label{eq:SHdecompVect}
\end{align}
(and similarly for $\vect{\xi}_\textrm{n}$), where $Y_{lm}(\theta,\phi)$ is the spherical harmonic function.  The oscillation equations then reduce to a set of linearly coupled ordinary differential equations in radius. In Appendix \ref{subsec:oscEq} we write down the form of these equations that we use in order to obtain numerical solutions.  As in \cite{Kantor:14}, we assume that the crust is a normal fluid.  In Appendix \ref{subsec:boundCond} we give the boundary conditions that we assume at the stellar center, at the core-crust interface (i.e., at the superfluid-normal fluid interface),  and at the stellar surface.  

In a normal fluid NS we can write the oscillation equations in the form of an eigenvalue problem
\begin{equation}
\mathcal{L}\left[\vect{\xi}\right]=\sigma^2\vect{\xi},
\end{equation}
where $\mathcal{L}\left[\vect{\xi}\right]$ is a linear operator representing the internal restoring forces that act on the Lagrangian displacement $\vect{\xi}(\vect{x},t)$.  The eigenmodes $\{(\sigma_a^2, \vect{\xi}_a)\}$ are those solutions that satisfy the boundary conditions, where
$a=\{n_a, l_a, m_a\}$ labels the three quantum numbers of each solution: the radial order $n_a$, the spherical degree $l_a$, and the azimuthal order $m_a$. Since the operator
$\mathcal{L}$ is Hermitian with respect to the inner product 
\begin{equation}
\left\langle \vect{\xi},\vect{\xi}'\right\rangle = 
\int \diff^3x \rho\, \vect{\xi}^\ast \cdot \vect{\xi}'
\end{equation}
(i.e., $\left\langle \vect{\xi}, \mathcal{L}\left[\vect{\xi}'\right]\right\rangle=\left\langle \mathcal{L}\left[\vect{\xi}\right], \vect{\xi}'\right\rangle$), the eigenmodes  form a complete, orthonormal basis (here the asterisk refers to complex conjugation). When considering normal fluid models, we normalize the modes such that
\begin{equation}
\sigma_a^2 \left\langle \vect{\xi}_a, \vect{\xi}_b\right\rangle  = E_0 \delta_{ab},
\end{equation}
where $E_0=GM^2/R$ is a characteristic energy scale of the NS.

While a normal fluid NS has a single displacement field $\vect{\xi}$, a superfluid NS has two distinct displacement fields $\vect{\xi}_\textrm{c}$ and $\vect{\xi}_\textrm{n}$ because there are two fluid components, the normal fluid and the superfluid.  The oscillation equations of a superfluid NS (eqs. \ref{eq:mass1}-\ref{eq:poisson}) therefore take the form
\begin{equation}
\mathcal{L}
\begin{bmatrix}
\vect{\xi}_+ \\
\vect{\xi}_-
\end{bmatrix}
=\sigma^2
\begin{bmatrix}
\vect{\xi}_+ \\
\vect{\xi}_-
\end{bmatrix},
\label{eq:free_sf_osc}
\end{equation}
where this linear operator $\mathcal{L}$ is different from that of the normal fluid case above (see Appendix \ref{subsec:oscEq}) and  
\begin{align}
\vect{\xi}_{+}  &= \frac{1}{\rho} (\rho_\textrm{c} \vect{\xi}_\textrm{c} + \rho_\textrm{n}\vect{\xi}_\textrm{n}), \label{eq:xiPlus}\\
\vect{\xi}_{-}  &=(1-\epsilon_\textrm{n} - \epsilon_\textrm{c})(\vect{\xi}_\textrm{c} - \vect{\xi}_\textrm{n}).  \label{eq:xiMinus}
\end{align}
The displacement $\vect{\xi}_{+}$ is the mass-averaged flow and the displacement $\vect{\xi}_{-}$ is proportional to the difference between the normal fluid flow and the superfluid flow.  For the tidal coupling problem, it proves convenient to express displacements in terms of $\vect{\xi}_{+}$ and $\vect{\xi}_{-}$ rather than $\vect{\xi}_\textrm{c}$ and $\vect{\xi}_\textrm{n}$. Note that although there is no direct force between the normal fluid and superfluid, they are nevertheless coupled locally through the equation of state (they are coupled even if entrainment is ignored; see discussion in \citealt{Prix:02}). As a result, both components oscillate at the same frequency $\sigma$.  The eigenmodes $\{(\sigma_a^2, \vect{\xi}_{a+}, \vect{\xi}_{a-})\}$ are those solutions that satisfy the boundary conditions given in Appendix \ref{subsec:boundCond}. In Appendix \ref{subsec:hermiticity} we show that the linear operator is Hermitian with respect to the inner product
\begin{equation}
\left\langle 
\begin{bmatrix}
 \vect{\xi}_{+} \\ \vect{\xi}_{-}
\end{bmatrix},
\begin{bmatrix}
\vect{\xi}_{+}' \\
\vect{\xi}_{-}' 
\end{bmatrix}\right\rangle
=\int \diff^3 x 
\begin{bmatrix}
 \vect{\xi}_{+}^\ast & \vect{\xi}_{-}^\ast
\end{bmatrix}
\begin{bmatrix}
\rho & 0 \\
0 & \tilde{\rho}
\end{bmatrix}
\begin{bmatrix}
\vect{\xi}_{+}' \\
\vect{\xi}_{-}' 
\end{bmatrix}
\label{eq:innerProduct}
\end{equation}
where 
\begin{equation}
\tilde{\rho}=\frac{\rho_\textrm{c}\rho_\textrm{n}}{(1-\epsilon_\textrm{n} - \epsilon_\textrm{c})\rho}.
\end{equation}
This result follows directly from the analysis in \cite{Lindblom:94} who, using somewhat different notation, showed that the linear operator satisfies a variational principle (see also \citealt{Andersson:01} and, for the case of a rotating NS, \citealt{Andersson:04}). The above integral reduces to the normal fluid case if we identify $\vect{\xi}_{+}  \to \vect{\xi}$ and $\vect{\xi}_{-} \to 0$, which allows us to evaluate it not only in the superfluid core but also in the normal fluid crust. We normalize the modes such that
\begin{equation}
\sigma_a^2\left\langle 
\begin{bmatrix}
 \vect{\xi}_{a+} \\ \vect{\xi}_{a-}
\end{bmatrix},
\begin{bmatrix}
\vect{\xi}_{b+} \\
\vect{\xi}_{b-} 
\end{bmatrix}\right\rangle
= E_0\delta_{ab}.
\label{eq:SForthonormal}
\end{equation}

\begin{figure*}
\hspace*{-0.5cm}  
\includegraphics[angle=0,scale=.48]{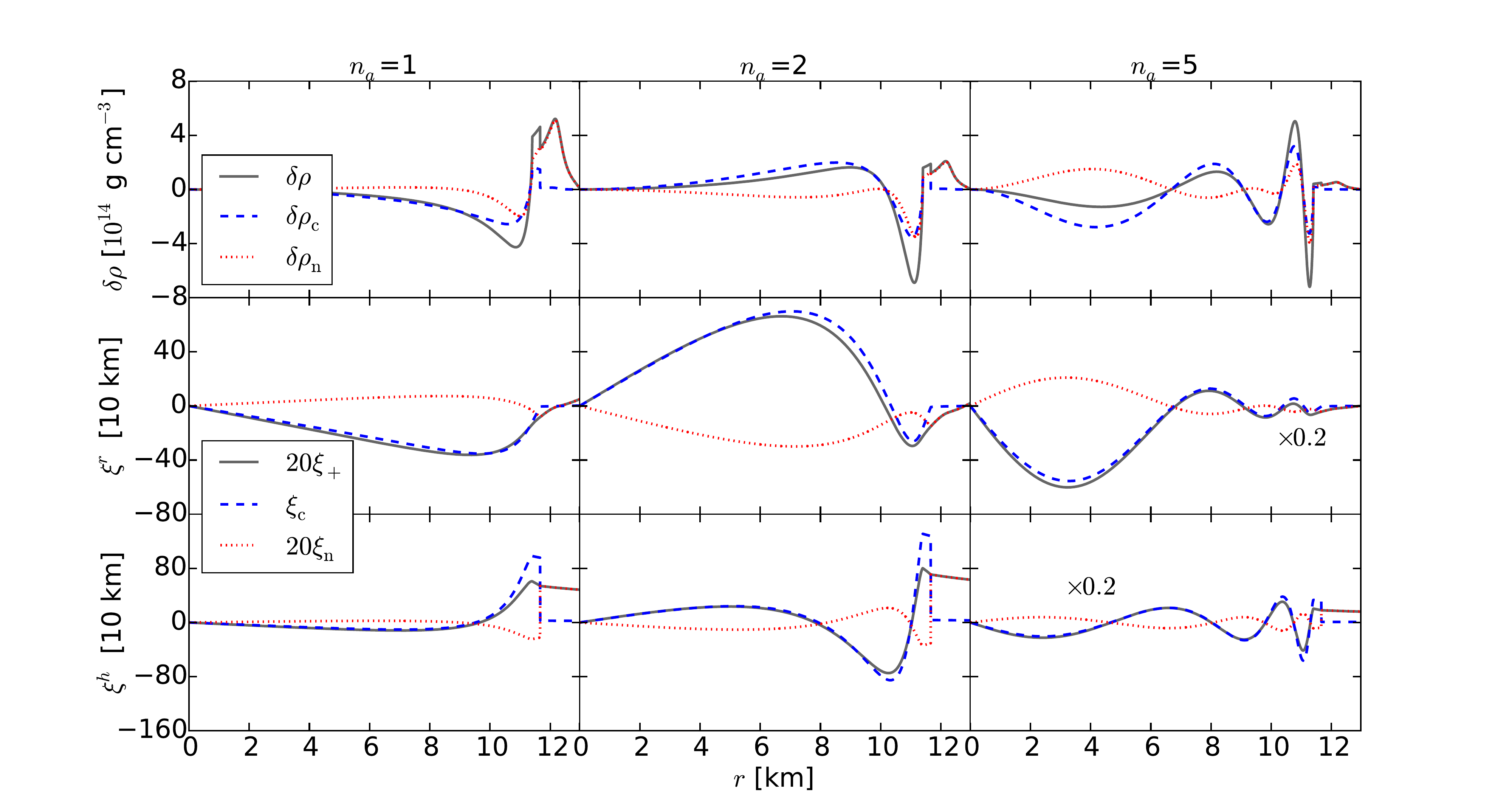}
\caption{Structure of the $l_a=2$, $n_a=(1,2,5)$ g modes (left, middle, and right panels, respectively) of our $1.4 M_\odot$ superfluid NS model with entrainment $m_\textrm{p}^\ast=0.8m_\textrm{N}$.   The upper panels show the total Eulerian density perturbation $\delta\rho=\delta\rho_\textrm{c}+\delta\rho_\textrm{n}$ (solid black line), $\delta\rho_\textrm{c}$ (dashed blue line), and $\delta\rho_\textrm{n}$ (dotted red line). The middle and lower panels show, respectively, the radial $\xi^r$ and horizontal $\xi^h$ components of the Lagrangian displacements corresponding to $\vect{\xi}_+$ (solid black line), $\vect{\xi}_\textrm{c}$ (dashed blue line), and $\vect{\xi}_\textrm{n}$ (dotted red line).  In order to plot all the displacements on the same scale, we multiply $\vect{\xi}_+$ and $\vect{\xi}_\textrm{n}$ by a factor of 20 and divide the $n_a=5$ displacements by a factor of 5. }
\label{fig:modeStruct}
\end{figure*}

In Fig. \ref{fig:modeStruct} we show the structure of three $l_a=2$ g modes ($n_a=1,\ 2,\ 5$) for the 1.4 $M_\odot$ superfluid  model with $m_\textrm{p}^\ast = 0.8 m_\textrm{N}$. In the top panel we plot the radial profile of the total  density perturbation $\delta\rho=\delta\rho_\textrm{c}+\delta\rho_\textrm{n}$ and that of the individual fluid components $\delta\rho_\textrm{c}$ and $\delta\rho_\textrm{n}$. In the bottom two panels we plot the radial and horizontal displacements of the two flows.  

There exists a discontinuity in the first derivative of $\delta \rho$ at $R_\mu$ because the muon gradient is discontinuous at $R_\mu$ in our model. There also exists a discontinuity in $\delta \rho$ at $R_{\rm cc}$ where we join the superfluid solution with the normal fluid solution. Nevertheless this discontinuity does not violate any physical principles.  In particular, it does not imply a discontinuous mass current since $\vect{\vel{v}} \delta \rho + \rho \delta \vect{\vel{v}}$ is still continuous:  the first term is always zero because the background velocity is zero, which suppresses the discontinuity in $\delta \rho$, and the second term is continuous by requiring continuity of the Lagrangian displacements (see Appendix \ref{subsec:boundCond}). 

For a given mode, the amplitude of $\vect{\xi}_\textrm{c}$ is significantly larger than $\vect{\xi}_\textrm{n}$ (in Fig.  \ref{fig:modeStruct} we multiply $\vect{\xi}_\textrm{n}$ by a factor of $20$ in order to plot it on a similar scale as $\vect{\xi}_\textrm{c}$).  This is because there is significantly less mass in the charged fluid elements (by a factor of $\simeq x_\textrm{p}$) and thus, for a given mode energy, $|\vect{\xi}_\textrm{c}|$ must be larger. We also find that $\xi^r_\textrm{c}$ and $\xi^r_\textrm{n}$ cross zero at slightly different locations (e.g., middle panel of Fig. \ref{fig:modeStruct}).  This effect is due to entrainment and was also observed by \cite{Prix:02} in the case of p-mode oscillations. Finally, we note that  in Fig. \ref{fig:modeStruct} the horizontal displacements $\xi^h$ are roughly twice as great as the radial displacements $\xi^r$, indicating the transverse nature of the g mode oscillations. 

\begin{figure}
\hspace*{-0.3cm}  
\includegraphics[angle=0,scale=.42]{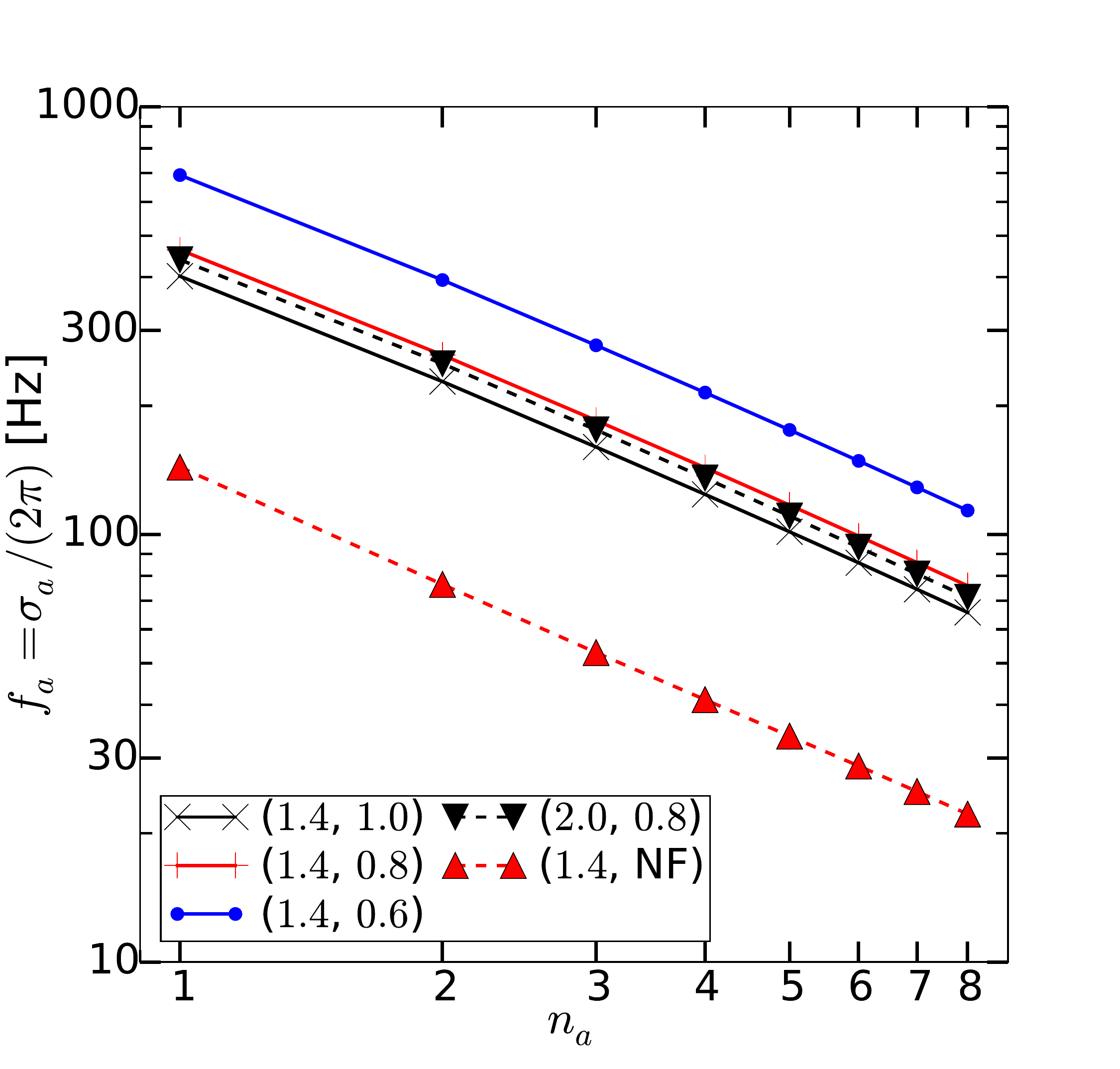}
\caption{Eigenfrequencies $f_a=\sigma_a/2\pi$ as a function of radial order $n_a$ for the first eight $l_a=2$ g modes.  In the legend, the first number represents the NS mass in units of $M_\odot$ and the second number represents $m_\textrm{p}^\ast$ in units of $m_\textrm{N}$ (or `NF' for the normal fluid case). We show results for four superfluid models:  $(M/M_\odot,\ m_\textrm{p}^\ast/m_\textrm{N}) = (1.4,\ 1.0)$, $(1.4,\ 0.8)$, $(1.4,\ 0.6)$, $(2.0,\ 0.8)$, and a normal fluid model with $M=1.4\ M_\odot$.  }
\label{fig:modeFreq}
\end{figure}

In Fig. \ref{fig:modeFreq} we show the eigenfrequencies $f_a=\sigma_a/2\pi$ of the first eight $l_a=2$ g modes for our various NS models.  Comparing the superfluid and normal fluid models, we see that the g mode spectra of the superfluid models are shifted to higher frequencies at a given $n_a$. This effect was also noted by \cite{Kantor:14} and \cite{Passamonti:16}.  The spectra shift because the buoyancy frequencies $\mathcal{N}$ are different in different models (see Section \ref{sec:buoyancy}); for high-order g modes  \citep{Aerts:10}
\begin{equation}
f_a \simeq \frac{\left[l_a(l_a+1)\right]^{1/2}}{2 \pi^2\,{\rm n}_a}\int N\diff \ln r.
\label{eq:gModePeriod}
\end{equation}
Indeed, we find that for even relatively low order $l_a=2$ g modes of the superfluid and normal fluid models,
\begin{equation}
f_a^{(\rm SF)}\simeq \frac{590}{n_a}\textrm{ Hz},
\hspace{0.6cm}\textrm{ and}\hspace{0.6cm}
f_a^{(\rm NF)}\simeq \frac{170}{n_a}\textrm{ Hz}.
\label{eq:fa_na_fit}
\end{equation}

The relations above are fits to the $1.4 M_\odot$ NS superfluid and normal fluid models, respectively; in the superfluid case we adopt an entrainment level of $m_\textrm{p}^\ast = 0.8 m_\textrm{N}$. We use these as our default models when providing numerical fits below. Among the superfluid models, increasing $m_\textrm{p}^\ast$ (that is, decreasing $\epsilon_\textrm{n}$) or increasing the NS mass decreases the eigenfrequencies slightly.  Equation (\ref{eq:fa_na_fit}) implies that in the frequency bandwidth of Advanced LIGO at full design sensitivity ($10\textrm{ Hz}-3000 \textrm{ Hz}$; \citealt{Harry:10}), a superfluid NS has $\approx 3$ times more $l=2$ g modes than a normal fluid NS.

\subsection{Tidal driving of modes}
\label{sec:homog}

We can account for tidal driving of the fluid by replacing $\delta \Phi$ in   equations (\ref{eq:mass1}-\ref{eq:poisson}) with $\delta \Phi + U$, where $U$ is the tidal potential.  In a spherical coordinate system $(r,\ \theta,\ \phi)$ centered on the primary, the tidal potential due to a companion of mass $M'$ is
\begin{equation}
U(r,\ \theta,\ \phi,\ t)=-GM'\sum_{l\ge2}\sum_{m=-l}^{l}\frac{W_{lm} r^l}{D^{l+1}(t)}Y_{lm}(\theta,\ \phi)e^{-im\psi(t)}.
\label{eq:tidal_potential}
\end{equation}
The orbit of the companion is oriented in the plane $(D(t),\ \pi/2,\ \psi(t))$, where $D(t)$ is the binary's orbital separation and $\psi(t)$ is the orbital phase.   The general expression for the coefficients $W_{lm}$ can be found in \citet{Press:77}; for the $l=2$ harmonic, which dominates at small $R/D$, $W_{20}=-\sqrt{\pi/5}$, $W_{2\pm2}=\sqrt{3\pi/10}$, and $W_{2\pm1}=0$. 
The superfluid oscillation equations with tidal driving now take the form
\begin{equation}
\left[\frac{\partial^2}{\partial t^2}+\mathcal{L}\right]
\begin{bmatrix}
\vect{\xi}_+ \\
\vect{\xi}_-
\end{bmatrix}
=
-\begin{bmatrix}
\vect{\nabla} U \\
0
\end{bmatrix}.
\label{eq:sf_forced}
\end{equation}
 The tidal acceleration $\vect{\nabla} U$ appears explicitly in the equation of the mass-averaged flow $\vect{\xi}_+$ but not the difference flow $\vect{\xi}_-$.  The normal fluid counterpart to equation (\ref{eq:sf_forced}) is recovered by identifying $\vect{\xi}_+ \rightarrow \vect{\xi}$ and $\vect{\xi}_- \rightarrow 0$ (see \citealt{Lai:94}).

\begin{figure*}
\includegraphics[angle=0,scale=.42]{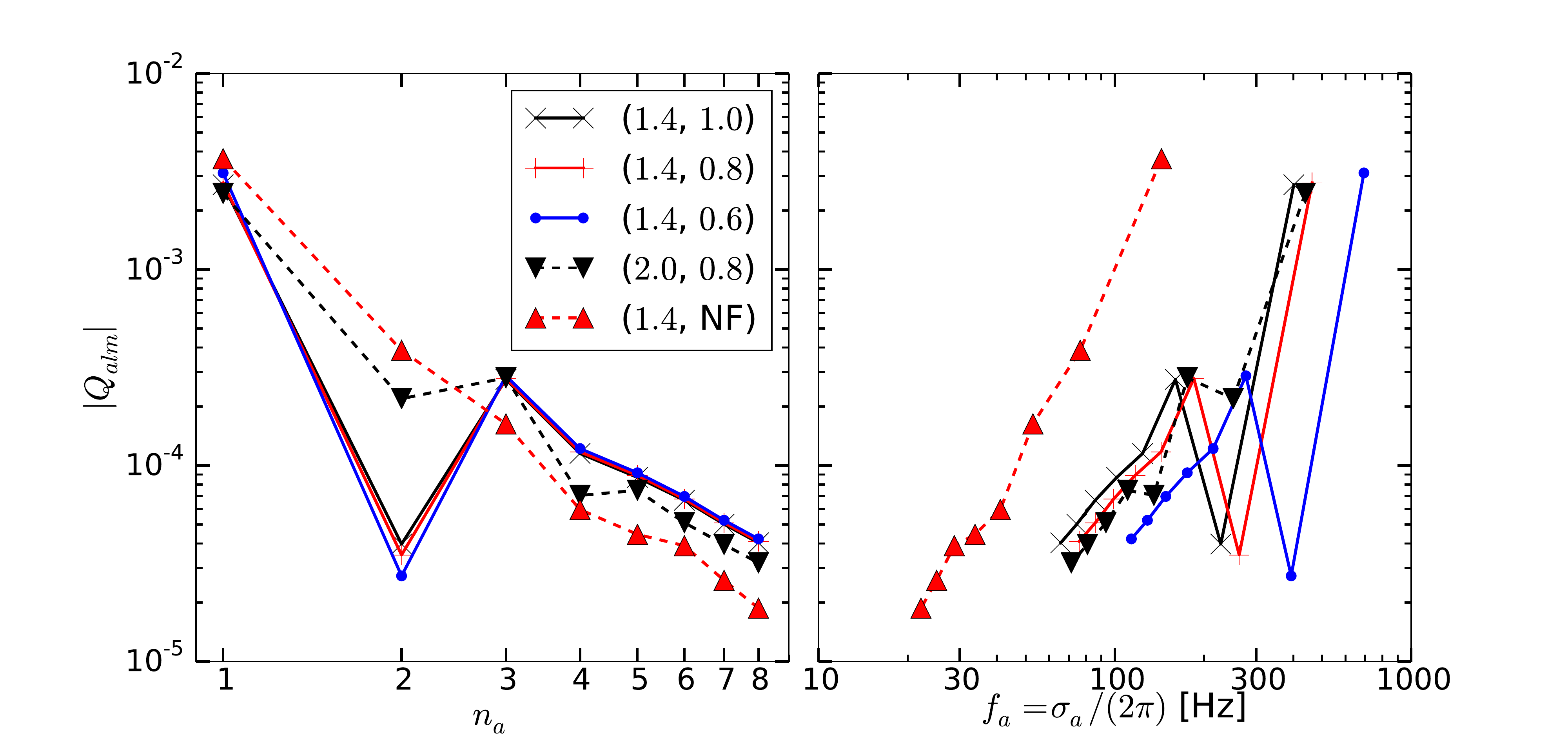}
\caption{Tidal coupling coefficient $|Q_{alm}|$ as a function of the $l_a=2$ g mode radial order $n_a$ (left panel) and eigenfrequency $f_a=\sigma_a/2\pi$ (right panel) for the same set of models as in Fig. \ref{fig:modeFreq}.}
\label{fig:coupCoeff}
\end{figure*}

Since the linear operator $\mathcal{L}$ is Hermitian (for both the superfluid and normal fluid; Appendix \ref{subsec:hermiticity}), the star's eigenmodes form an orthonormal basis. This allows us to expand the displacements as 
\begin{equation} 
\begin{bmatrix}
\vect{\xi}_+ (\vect{x},t)\\
\vect{\xi}_-(\vect{x},t)
\end{bmatrix}
=\sum_a b_a(t)
\begin{bmatrix}
\vect{\xi}_{a+} (\vect{x})\\
\vect{\xi}_{a-}(\vect{x})
\end{bmatrix},
\label{eq:mode_decomp}
\end{equation}
where $b_a(t)$ is the time-dependent, dimensionless amplitude of mode $a$. Given our eigenmode normalization (eq. \ref{eq:SForthonormal}), a mode with amplitude $|b_a|=1$ has energy $E_0$. Equation (\ref{eq:sf_forced}) can then be written as a set of linear amplitude equations for each mode:
\begin{equation}
\ddot{b}_a + \sigma_a^2 b_a =\sigma_a^2 U_a(t)
\label{eq:amp_eqn}
\end{equation}
where the tidal driving coefficient (cf. \citealt{Weinberg:12})
\begin{align}
U_a(t) &= -\frac{1}{E_0}\int \diff^3x \rho \, \vect{\xi}_{a+}^\ast \cdot \vect{\nabla} U\\
&= \frac{M'}{M}\sum_{lm} W_{lm} Q_{alm} \left(\frac{R}{D(t)}\right)^{l+1} e^{-im\psi(t)}.
\label{eq:tide_coefficient}
\end{align}
The second equality follows from equation (\ref{eq:tidal_potential}) and defines the time-independent, dimensionless tidal coupling coefficient (sometimes referred to as the tidal overlap integral)
\begin{equation}
Q_{alm} =\frac{1}{MR^l}\int \diff^3 x \rho \, \vect{\xi}_{a+}^\ast \cdot\vect{\nabla}\left(r^l Y_{lm}\right),
\label{eq:Qform_orig}
\end{equation}
where in the subscripts $a=\{n_a,\ l_a,\ m_a\}$ denotes a specific eigenmode of the NS and $lm$ denotes a specific harmonic of the tidal potential. By angular momentum conservation, $Q_{alm}$ is non-zero only if $l_a=l$ and $m_a=m$. Using equations (\ref{eq:mass1}), (\ref{eq:mass2}), and (\ref{eq:poisson}) and integrating by parts we can alternatively express the tidal coupling coefficient as
\begin{equation}
Q_{alm} =\frac{1}{MR^l}\int \diff r r^{l+2} \delta \rho_{a} = -\frac{2l+1}{4\pi}\frac{\delta \Phi_a(R)}{GM/R},
\label{eq:coupCoefForms}
\end{equation}
where $\delta \rho_a=\delta\rho_{\textrm{c},a}+\delta\rho_{\textrm{n},a}$ is the total perturbed density due to mode $a$ and $\delta \Phi_a(R)$ is the mode's perturbation to the gravitational potential at the stellar surface.

In Fig. \ref{fig:coupCoeff} we show $|Q_{alm}|$ as a function of the radial order $n_a$ and eigenfrequency $f_a=\sigma_a/2\pi$ of the $l_a=2$ g modes for our various NS models. The most obvious feature is that smaller $n_a$ tend to have larger $|Q_{alm}|$ (with the exception of the $n_a=2$ mode of our $1.4 M_\odot$ superfluid models, which has an anomalously small $|Q_{alm}|$).  This is because the tide is a long wavelength perturbation and it couples best to modes whose wavelengths are likewise long.  For a given $n_a$, we find that the different models all have similar $|Q_{alm}|$; there is only a weak dependence on whether the NS is superfluid, the level of entrainment $m_\textrm{p}^\ast$, and the NS mass.  Since the superfluid g mode spectrum is shifted to higher frequencies (i.e., $f_a$ is larger at a given $n_a$),  at a given $f_a$ the normal fluid models have a significantly larger $|Q_{alm}|$.  In particular, based on our numerical calculations of $|Q_{alm}|$, we find that for the $1.4 M_\odot$ superfluid models (neglecting the anomalous $n_a=2$ mode) and normal fluid models, respectively, 
\begin{align}
\left|Q_{alm}^{(\rm SF)}\right|
&\simeq 2.6\times 10^{-3} n_a^{-2}
\simeq 7.6\times 10^{-5} f_{a,100}^2,
\label{eq:Qsf}\\
&\nonumber\\
\left|Q_{alm}^{(\rm NF)}\right| 
&\simeq 3.5\times 10^{-3} n_a^{-5/2}
\simeq 9.3\times 10^{-4} f_{a,100}^{5/2},
\label{eq:Qnf}
\end{align}
where $f_{a,100}=f_a/100\textrm{ Hz}$. The oscillatory nature of the g modes can make calculating $Q_{alm}$ subject to numerical error \citep{Reisenegger:94, Reisenegger:94b, Weinberg:12}.  In Appendix \ref{sec:numErr} we carry out numerical tests that show that our calculations of $Q_{alm}$ have only a $\sim 1$ per cent error.
 
 \section{Results}
\label{sec:energyPhase}

Using the formalism described in the previous section, we now evaluate the resonant tidal excitation of g modes in coalescing superfluid NS binaries.  Our analysis is similar to that of \cite{Lai:94} and \cite{Reisenegger:94} who studied this problem for normal fluid NSs. In Section \ref{sec:energy_transfer} we calculate the energy transferred to the NS from the orbit due to the resonant tidal interactions.  In Section \ref{sec:phase_error} we calculate the resulting GW phase error relative to the point mass estimate.

\subsection{Tidal energy transfer}
\label{sec:energy_transfer} 

As the NS inspirals due to the emission of gravitational radiation, the tidal driving sweeps through resonances with individual g modes. The dynamics, which are similar to that of a linearly driven oscillator whose driving frequency and forcing strength increase with time, is determined by the amplitude equation (\ref{eq:amp_eqn}).  Focusing on resonant driving by the dominant $l=2, m=2$ tidal harmonic, we have
\begin{equation}
\ddot{b}_{a} +\sigma_{a}^2b_a = \sigma_a^2 W_{22} Q_{a22} \left(\frac{M'}{M}\right) \left(\frac{R}{D(t)}\right)^3 e^{-2i\psi(t)}.  
\label{eq:amp_eq_lm2}
\end{equation}
Since linear tidal interactions have a small overall effect on the inspiral,  we can use the quadrupole formula for the rate of orbital decay of two point masses, i.e.,
\begin{align}
\label{eq:sep}
\dot{D}&=-\frac{64G^3}{5c^5}\frac{MM'(M+M')}{D^3}, \\
\label{eq:orbPhase}
\dot{\psi}&=\left[\frac{G(M+M')}{D^3}\right]^{1/2}.
\end{align}
As $D(t)$ decreases and the orbital frequency $\Omega(t)=\dot{\psi}$ increases, $l_a=2$ g modes with $\sigma_a\simeq 2\Omega$ temporarily undergo resonant tidal driving. Post-resonance, the g modes oscillate at nearly their natural frequency $\sigma_a$ \citep{Lai:94}. 

We do not include linear damping in equation (\ref{eq:amp_eq_lm2}) because it has a negligible effect on the peak amplitudes reached by the low order modes we consider. It therefore does not affect the tidal energy transfer or phase error.  Damping does heat the neutron star by thermalizing a portion of the mode energy. Nonetheless, as we show later in this section, the core is only heated to $T \sim 10^7\textrm{ K}$, which is too small to significantly modify the g modes of a superfluid NS \citep{Kantor:14, Passamonti:16}. 

\begin{figure*}
\hspace*{-1cm}  
\includegraphics[angle=0,scale=.42]{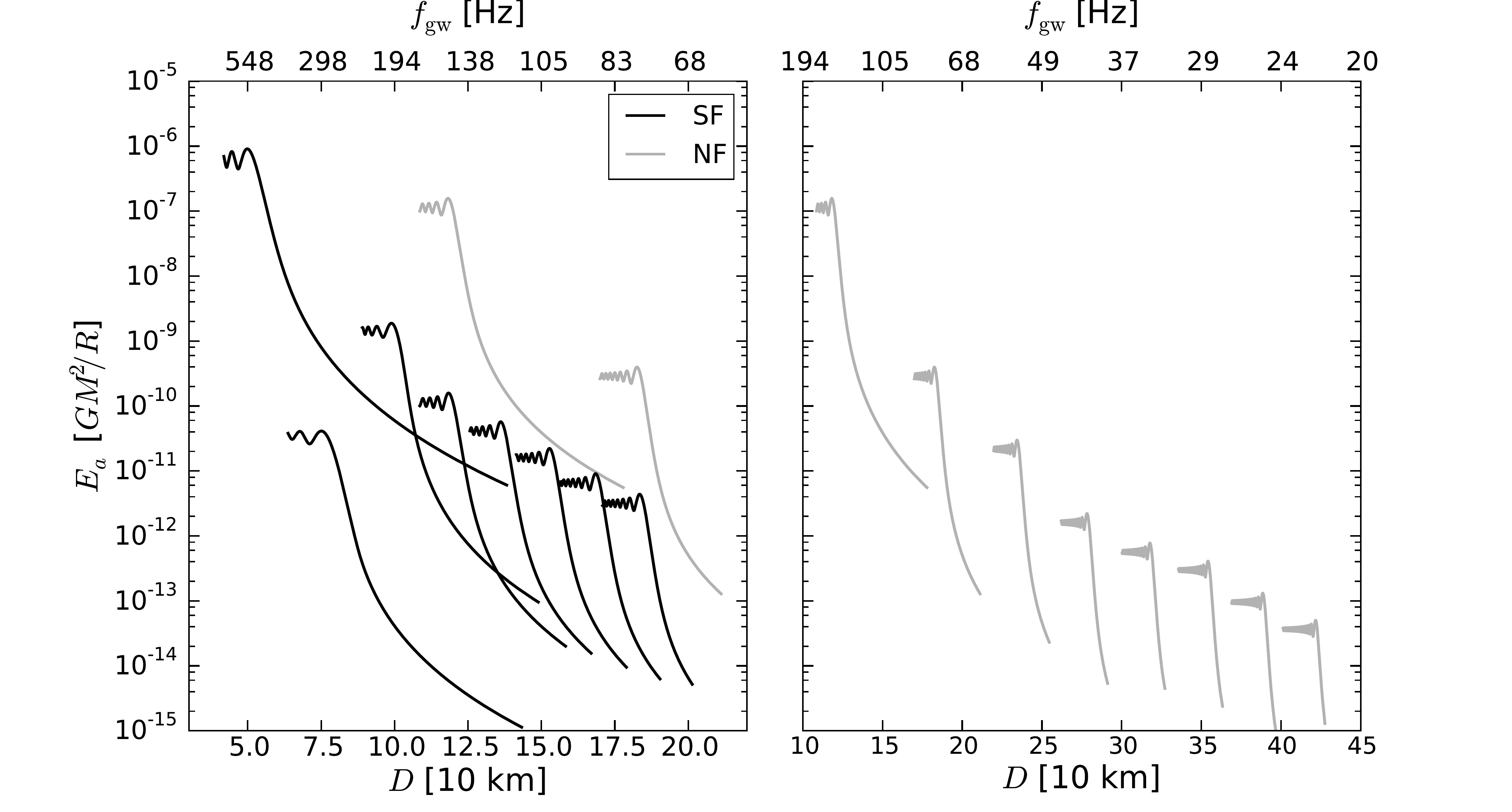}
\caption{Evolution of the mode energy $E_a$ (in units of $E_0=GM^2/R$) due to the resonant tidal driving of $l_a=2$ g modes during an equal mass ($M=M'=1.4M_\odot$) binary NS inspiral. The bottom axes gives the orbital separation $D$ and the top axes give the gravitational wave frequency $f_{\rm gw}$.  The left panel shows the $n_a=\{1,2,3,\ldots 8\}$ g modes of the superfluid model with $M=1.4\ M_\odot, m_\textrm{p}^\ast=0.8m_\textrm{N}$ (black lines) and the $n_a=\{1,2\}$ g modes of the normal fluid model with $M=1.4\ M_\odot$ (grey lines).  The right panel shows the $n_a=\{1,2,3,\ldots 8\}$ g modes of the normal fluid model (the $n_a=\{1,2\}$ modes are plotted in both panels).  Note the different range of $D$ plotted in the two panels. For clarity, we only show a mode's evolution near its resonant excitation.}
\label{fig:Etide}
\end{figure*}

In order to determine the evolution of the mode amplitudes $b_a(t)$, we solve equations (\ref{eq:amp_eq_lm2}), (\ref{eq:sep}), and (\ref{eq:orbPhase}) for the set of g modes described in Section \ref{sec:Formalism}. 
For each mode we initialize the equations following the discussion in \cite{Lai:94}, and then numerically integrate them forward in time. In Fig. \ref{fig:Etide} we show the mode energy $E_{a}(t)=2 |b_a|^2E_0$ as a function of orbital separation $D(t)$ for the low order $(l_a=2,\ m_a=\pm 2)$ g modes that are resonantly excited during the latter stages of inspiral ($f_{\rm gw}\gtrsim 30\textrm{ Hz}$). For conciseness, we have used a single letter $a$ in the subscript of mode energy to represent the total contribution of both the $m_a=2$ and $m_a=-2$ modes, and thus a factor of 2 has been included since each mode contributes equally. We will use this convention in all our results described below. In the left panel we show $E_a(t)$ for our superfluid NS model with $M=M'=1.4 M_\odot$ and $m_\textrm{p}^\ast=0.8 m_\textrm{N}$. In the right panel we show $E_a(t)$ for the normal fluid NS model with $M=M'=1.4 M_\odot$.  Note that the horizontal scale is different in the two panels. 

Because the superfluid model has more high-frequency g modes (see Fig. \ref{fig:modeFreq}), it admits eight resonantly excited g modes for $D(t)<200\textrm{km}$ compared to only two for the normal fluid model. The lowest order superfluid g mode is excited later in the inspiral than the normal fluid one (compare the black and grey curves in the left panel of Fig. \ref{fig:Etide}). On the other hand, at orbital separations where both models have resonances, the modes of the normal fluid model are excited to a significantly larger maximum energy $E_{a, \rm max}$. For example, at $D\simeq 120\textrm{ km}$, the $n_a=1$  mode of the normal fluid model undergoes resonant driving up to $E_{a,\rm max} \approx 10^{-7} E_0$ while the $n_a=4$ mode of the superfluid model undergoes resonant driving up to only $E_{a,\rm max} \approx10^{-10} E_0$. This difference is due to the superfluid model's smaller tidal coupling coefficient $|Q_{a22}|$ at a given $f_a$ (see Section \ref{sec:homog}).

While the numerical calculations provide the full mode amplitude evolution, we can estimate the post-resonance mode energy $E_{a, \rm max}$ by solving equation (\ref{eq:amp_eq_lm2}) using the stationary-phase approximation.  Following the approach described in \citeauthor{Lai:94} (1994; see also \citealt{Reisenegger:94}), this gives 
\begin{equation}
E_{a,\rm max} \simeq \frac{\pi^2}{1024} k \left(\frac{GM}{Rc^2}\right)^{-5/2}  \left(\frac{\sigma_a}{\omega_0}\right)^{7/3}\sum_{m=\pm2}Q_{a2m}^2 E_0,
\label{eq:SPApproxEnergy}
\end{equation}
where $k=q[2/(1+q)]^{5/3}$, $q=M'/M$ is the mass ratio of the binary, and $\omega_0=(GM/R^3)^{1/2}$ is the NS dynamical frequency.  The expression matches equation (6.11)  in \cite{Lai:94} except that we use a different convention for normalizing the eigenfunctions.  Using our analytic fits to $Q_{a22}$ given by equations (\ref{eq:Qsf}) and (\ref{eq:Qnf}) and the values of $M$ and $R$ given in table \ref{tab:bgConfig}, we find that for the $M=1.4M_\odot$ superfluid and normal fluid  models, respectively,
\begin{align}
E_{a,\rm max}^{(\rm SF)} 
&\simeq 1.0\times10^{-6}k\, n_a^{-19/3}\, E_0 
\label{eq:Ea_sf_na}
\simeq 2 \times10^{-11} k\, f_{a,100}^{19/3} \,E_0, \\
&\nonumber \\
E_{a,\rm max}^{(\rm NF)} 
&\simeq 1.2\times10^{-7}k\, n_a^{-22/3} \, E_0 
\label{eq:Ea_nf_na}
\simeq 3.0 \times10^{-9} k\, f_{a,100}^{22/3}\, E_0,  
\end{align}
where  we used equation (\ref{eq:fa_na_fit}) to express the energies in terms of both $n_a$ and $f_a$.  Comparing this with the fully numerical results shown in Fig. \ref{fig:Etide}, we find that the stationary-phase approximation gives a good match to the superfluid energy $E_{a, \textrm{max}}^{(\textrm{SF})}$ but slightly underestimates the normal fluid case by $\approx 25$ per cent.  At a given frequency, we find that $E_{a,\rm max}$ of both $M=2.0M_\odot$ superfluid and normal fluid models are both about $3$ times smaller than $E_{a,\rm max}$ of the $M=1.4M_\odot$ models. 
 
In order to calculate the total energy transfer $E_{\rm trans}$ from the orbit to all the $l=2$ g modes, we can sum over $n_a$ using equations (\ref{eq:Ea_sf_na}) and (\ref{eq:Ea_nf_na}). This gives
\begin{equation}
\label{eq:Etrans}
E_{\rm trans}^{(\rm SF)} \simeq 1.0\times 10^{-6} k\, E_0
\hspace{0.3cm} \textrm{and} \hspace{0.3cm}
E_{\rm trans}^{(\rm NF)} \simeq 1.2\times 10^{-7} k\, E_0.
\end{equation}
Thus, a superfluid NS absorbs $\simeq 10$ times more orbital energy by the time the NS merges. The sums over $n_a$, which formally are given by the Riemann zeta function $\zeta(19/3)\simeq \zeta(22/3)\simeq 1.0$, are strongly dominated by the $n_a=1$ mode. That is, most of the energy transfer occurs during the excitation of the lowest order g mode. This result is a consequence of two  effects: the tidal coupling coefficient $|Q_{a22}|$ is largest for low-order modes (see Section \ref{sec:homog}), and the amplitude of the tide $(M'/M)(R/D)^3$ is largest at small $D$, which is when the low-order (i.e., high $f_a$) modes are resonantly excited.  The influence of these two effects is only partially mitigated by the shorter decay timescales at small $D$, which reduces the duration of the resonant driving compared to higher-order modes. 

Following \cite{Lai:94}, viscous dissipation of the resonant g modes heats the neutron star by an amount
\begin{equation}
E_\textrm{visc}\simeq-2\int_{D_a}^{D_\textrm{merg}} \frac{\diff D}{D} t_D \gamma_a E_a,
\end{equation}
where $D_a$ is the orbital separation at which the mode $a$ becomes resonant, $D_\textrm{merg}$ is the separation before the merger (taken to be $3R$), $t_D=|D/\dot{D}|$ is the orbital decay time, and $\gamma_a$ is the mode's damping rate. We neglect the small amount of viscous dissipation of modes prior to their resonant excitation. Comparing the heating in the superfluid case relative to the normal fluid case, we find
\begin{equation}
\frac{E_\textrm{visc}^\textrm{(SF)}}{E_\textrm{visc}^\textrm{(SF)}} \simeq 0.3 \frac{\gamma_1^\textrm{(SF)}}{\gamma_1^\textrm{(NF)}},
\label{eq:viscHeating}
\end{equation}
where $\gamma_1$ represents the damping rate of the first g mode, which we expect to dominate the heating (although higher order modes have larger $\gamma_a$ and more time to heat the NS prior to the merger, they contribute less to the heating because their $E_{a, \rm max}$ is much smaller).  
Following \cite{Lai:94}, if we assume that the viscosity is dominated by electron-electron scattering and that the heat content is dominated by the electrons, then the superfluid NS is heated to $T\sim 10^7\textrm{ K}$.\footnote{Our estimate of the heating differs from that of \cite{Lai:94} in two ways.  First, since we are considering a superfluid NS rather than a normal fluid NS, we assume that the main thermal content is due to the electrons rather than the neutrons (see footnote 9 in \citealt{Lai:94}).  This increases the resulting temperature by a factor of $\approx 2$. Second, we correct a typo in \citeauthor{Lai:94}'s expressions for the damping rates which for $l=|m|=2$ modes decreases the rates by a factor of 24 (see footnote 14 in \citealt{Weinberg:13}).\label{foot:Tcorrection}}   Such temperatures are too small to significantly modify the g modes relative to the zero-temperature superfluid model we have adopted in our calculation (see, e.g., Fig. 4 in \citealt{Kantor:14} and \citealt{Passamonti:16}). 

\subsection{Phase shift of the gravitational waveform}
\label{sec:phase_error}

\begin{figure*}
\includegraphics[angle=0,scale=.42]{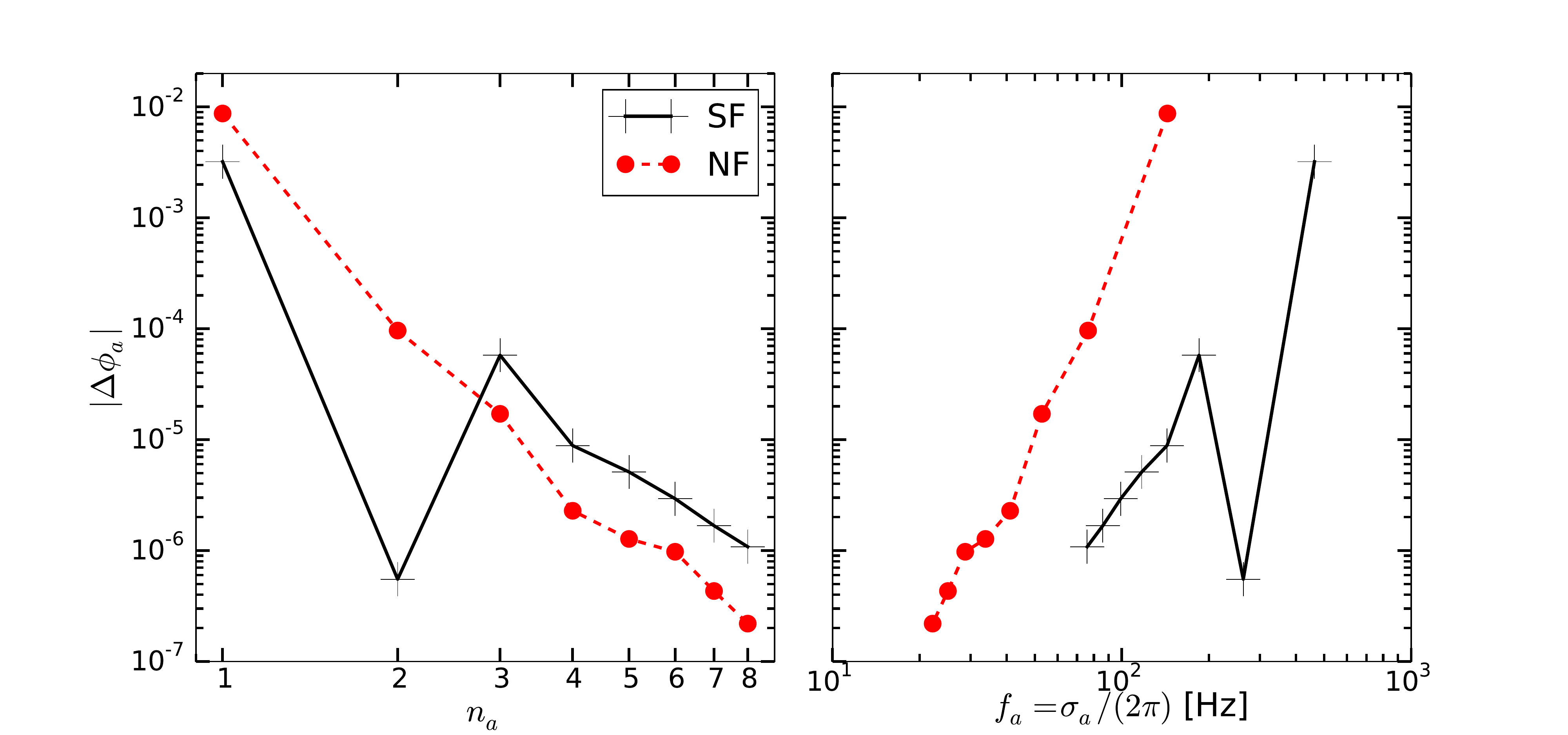}
\caption{Phase shift of the gravitational waveform $\Delta \phi_a$ due to the resonant tidal excitation of individual $l=2$ g modes. The left panel shows $\Delta \phi_a$ as a function of the radial order $n_a$ and the right panel as a function of the mode's eigenfrequency $f_a$. Solid black lines correspond to the $M=1.4 M_\odot$ superfluid NS model with an entrainment level $m_\textrm{p}^\ast = 0.8m_\textrm{N}$. For comparison, dashed red lines show the results for the $M=1.4 M_\odot$ normal fluid NS model.}
\label{fig:delN}
\end{figure*}

Given the resonant energy $E_{a, \rm max}$, the phase shift of the gravitational waveform $\Delta \phi_a$ due to each excited mode is given approximately by \citep{Lai:94}
\begin{equation}
\Delta \phi_a \simeq -4\pi\frac{t_D}{t_{\rm orb}}\frac{E_{a, \rm max}}{\left|E_{\rm orb}\right|},
\end{equation}
where $t_{\rm orb}=2\pi/\Omega$ is the orbital period and $E_{\rm orb}=-GMM'/2D$ is the orbital energy (both evaluated at the mode's resonance). Because the modes remove energy from the orbit, the tidal interaction accelerates the rate of orbital decay and thus $\Delta \phi_a < 0$. Using the expression for $E_{a, \rm max}$ based on the  stationary-phase approximation (eq. \ref{eq:SPApproxEnergy};  note that the contributions from both $m=\pm2$ modes are included), we find 
\begin{equation}
\Delta \phi_a=-\frac{5\pi^2}{2048}k'\left(\frac{GM}{Rc^2}\right)^{-5}\sum_{m=\pm2}|Q_{a2m}|^2,
\label{eq:dphia_genl}
\end{equation}
where $k'=2/[q(1+q)]$. Note that in our normalization,  $\Delta \phi_a$ depends on frequency only through $|Q_{a2m}|$.  Using equations (\ref{eq:Ea_sf_na}) and (\ref{eq:Ea_nf_na}), we find
\begin{align}
\Delta \phi_a^{(\rm SF)} 
&\simeq -3\times10^{-3} k' n_a^{-4}
\simeq -3\times10^{-7}k' f_{a,100}^4,
\label{eq:dphia_SF}\\
\Delta \phi_a^{(\rm NF)} 
&\simeq -7 \times10^{-3} k' n_a^{-5}
\simeq -4 \times10^{-4} k' f_{a,100}^5.
\label{eq:dphia_NF}
\end{align}
These analytic estimates of the phase error are in good agreement (to within $\simeq 25\%$) with the numerical results shown in Fig. \ref{fig:delN}.

As in the $E_{\rm trans}$ calculation of Section \ref{sec:energy_transfer}, we can sum over $n_a$ and $m_a$ to get the total phase error $\Delta \phi$ due to the excitation of all the $l=2$ g modes.  This gives
\begin{align}
\Delta \phi^{(\rm SF)} &\simeq -4\times10^{-3}k', \\
\Delta \phi^{(\rm NF)} &\simeq -7\times10^{-3}k'. 
\end{align}
As with $E_{\rm trans}$,  the strong scaling with $n_a$ in equations (\ref{eq:dphia_SF}) and  (\ref{eq:dphia_NF}) implies that the phase error is almost completely dominated by the resonant excitation of the lowest order modes.

Although  each g mode in a superfluid NS is, compared to a normal fluid NS, excited to a much greater energy [$\simeq 10$ times larger for the lowest order mode; see equations (\ref{eq:Ea_sf_na}) and (\ref{eq:Ea_nf_na})], it is excited later in the inspiral when the orbital decay is faster. These two effect cancel and therefore $\Delta \phi_a$ depends only on the tidal coupling strength $|Q_{alm}|$ [equation (\ref{eq:dphia_genl})]. For a given $n_a$, $|Q_{alm}|$ is insensitive to whether the NS is superfluid [equations (\ref{eq:Qsf}) and (\ref{eq:Qnf})] and, as a result, superfluid and normal fluid NSs have similar dynamical tide-induced GW phase shifts.

\section{CONCLUSIONS}\label{sec:conclusion}

We studied the dynamical tide in  coalescing superfluid NS binaries.  We considered NSs with an npe$\mu$ composition for different stellar masses ($M=1.4 M_\odot$ and $2 M_\odot$) and levels of entrainment (as quantified by the proton effective mass $m_p^\ast$).  Although we did not account for general relativistic effects in our calculations, this simplification is unlikely to influence the qualitative conclusions of our study. In all of our superfluid NS models, we found that the spectrum of the $l=2$ g modes is shifted to higher frequencies compared to a normal fluid NS. As a result, we showed that many more modes undergo resonant excitation during the latter stages of binary inspiral.  By calculating the mode coupling strength and integrating the time-dependent mode amplitude equations as the binary sweeps up in frequency, we found that the total energy transfer from the orbit to the oscillations is $\simeq 10$ times larger than the normal fluid case. However, because the energy transfer is dominated by the highest frequency modes, it occurs later in the inspiral when the orbital decay is faster.  As a result, the impact of tidal interactions on the GW signal is comparable for a superfluid and normal fluid NS.  In particular, the magnitude of the GW phase shift in both cases is $\simeq \textrm{ a few}\times10^{-3}\textrm{ radian}$.  Such a phase shift is at least two orders of magnitude too small to be detected by the current generation of GW detectors (see, e.g., \citealt{Cutler:94}).

Our analysis did not account for hyperons, which are expected to appear at high core densities ($\sim 7 \times 10^{14} \textrm{ g cm}^{-3}$; see, e.g., \citealt{Bednarek:12, Weissenborn:12, Gusakov:14}). As \cite{Dommes:16} point out, gradients in the hyperon fraction might also be a source of buoyancy in  superfluid NSs. While the direct Urca process involving hyperons (see review by \citealt{Yakovlev:01})  may be fast enough compared to the g mode oscillation period to break the assumption of frozen composition, and/or the hyperons may be superfluid themselves \citep{Takatsuka:06, Wang:10}, the case studied by \cite{Dommes:16} nonetheless shows that there can exist additional g modes in hyperonic NSs. In particular, hyperons produce an additional peak in the Brunt-V{\"a}is{\"a}l{\"a} frequency profile, one that occurs much deeper in the core than the peak due to the muon-to-electron gradient (see Fig. 6 in \citealt{Dommes:16}).  This will modify the properties of the g modes calculated here and it is not clear to what extent this might alter the conclusions of our analysis. We plan to address this problem in the future. 

We also did not account for NS rotation.  Studies that have find that rapid rotation can lead to significantly larger tide-induced phase shifts
\citep{Ho:99, Lai:06, Flanagan:07}.  However,  even though these studies all assume normal fluid NSs, the modes that are responsible for the largest phase shifts are f-modes, r-modes, and inertial modes.  Such modes are unlikely to be significantly modified by superfluid effects (e.g., \citealt{Lee:95,Passamonti:09}). 

It has been suggested that the tide in coalescing NS binaries becomes unstable to nonlinear fluid effects at relatively low GW frequencies ($\approx 50\textrm{ Hz}$; \citealt{Weinberg:13, Venumadhav:14, Weinberg:16}). Although these studies assume a normal fluid NS, the nonlinear effects involve non-resonant, low frequency g modes and such modes still exist in superfluid NSs.  However,  it is not clear to what extent superfluidity might alter the growth rate and saturation of the instability.  It would therefore be interesting to extend these studies to superfluid NSs. 

\section*{Acknowledgements}

The authors thank Jocelyn Read, Reed Essick, and the referee for detailed and valuable comments. This work is supported in part by NASA  ATP grant NNX14AB40G. HY is also supported in part by the National Science Foundation and the LIGO Laboratory. LIGO was constructed by the California Institute of Technology and Massachusetts Institute of Technology with funding from the National Science Foundation and operates under cooperative agreement PHY-0757058.

\bibliographystyle{mnras}
\bibliography{ref}{}

\appendix

\section{THERMODYNAMIC RELATIONS AND SUPERFLUID ENTRAINMENT}\label{sec:Thermodyn}
In this appendix we present the thermodynamic relations that we use in our study. In Section \ref{subsec:basicQuan} we give the expressions that we use in order to calculate the background quantities (such as density and pressure). In Section \ref{subsec:Entrainment} we describe our implementation of the entrainment effect and provide the connection between our notation and that used in previous studies.

\subsection{Background quantities}\label{subsec:basicQuan}
We model the superfluid neutron star as a zero-temperature system consisting of two fluids: the superfluid neutrons (denoted by subscript n for `neutrons') and a normal fluid mixture of  protons, electrons, and muons whose abundances are linked through charge neutrality (denoted by subscript c for `charged'). According to the thermodynamic identity, the total energy density $\varepsilon_\textrm{tot}$  satisfies
\begin{equation}
\diff\varepsilon_\textrm{tot} = \sum_{j=\textrm{npe}\mu} \mu_j \diff n_j + \alpha \diff \vel{v}_\textrm{r}^2,
\label{eq:thermoDynamicID}
\end{equation}
where $n_j$ and $\mu_j$ are the number density and chemical potential of particle species $j$ ($=$n, p, e, $\mu$), $\vect{\vel{v}}_\textrm{r} = \vect{\vel{v}}_\textrm{c} - \vect{\vel{v}}_\textrm{n}$ is the relative velocity between the charged and neutron flows, and $\alpha$ is a function representing the entrainment effect. Since the relative velocity between the two flows is small (and zero for the background model we consider here), we can separate the entrainment part from the bulk motion and write
\begin{equation}
\varepsilon_{\rm tot} = \varepsilon + \alpha \vel{v}_\textrm{r}^2,
\end{equation}
where the bulk energy density $\varepsilon$ can be represented as a sum of the baryonic and leptonic contributions
\begin{equation}
\varepsilon = \left(n_\textrm{n} +n_\textrm{p}\right) \left[m_\textrm{N}c^2 + E_\text{nuc}(n_\textrm{n},\ n_\textrm{p})\right]  + T_\textrm{e} +  T_\mu.
\end{equation}
Here $m_\textrm{N}$ is the nucleon rest mass, $E_\text{nuc}$ is the interaction energy per baryon given by the nuclear equation of state, and $T_\textrm{e}$ and $T_\mu$ are the total energy of the electrons and muons, respectively. We use the SLy4 nuclear equation of state with $E_\textrm{nuc}$ given by equation (3.18) in \citet{Chabanat:97}. We assume the leptons are described by a zero-temperature, relativistic free Fermi gas with
\begin{align}
&T_\textrm{e} = \frac{3}{4}\hbar c(3\pi^2)^{1/3}(n_\textrm{e})^{4/3}, \\
&T_\mu = \frac{m_\mu^4 c^5}{\hbar^3}\frac{1}{8\pi^2}\Big\{x\left(1+x^2\right)^{1/2}\left(1+2x^2\right)\nonumber \\
&\hspace{0.5cm}-\ln \left[x+\left(1+x^2\right)^{1/2}\right]\Big\},
\end{align}
where $x = p_F/m_\mu c = \hbar (3n\pi^2n_\mu)^{1/3}/m_\mu c$.  It is worth noting that to fully parameterize the bulk energy density $\varepsilon$ of the npe$\mu$ NS under the constraint of charge neutrality, we need three independent variables (for example, $n_\textrm{n}$, $n_\textrm{e}$ and $n_\mu$, with $n_\textrm{p}=n_\textrm{e}+n_\mu$ by charge neutrality; cf. equation \ref{eq:thermoDynamicID}). This is fundamentally different from the npe NS (studied by, e.g., \citealt{Lee:95}, \citealt{Andersson:01} and \citealt{Prix:02}), which requires only two independent variables.

The chemical potential for each species is given by
\begin{align}
&\mu_\textrm{n} = m_\textrm{N}c^2 +  E_\text{nuc} + n\frac{\partial E_\text{nuc}}{\partial n_\textrm{n}}, \\
&\mu_\textrm{p} = m_\textrm{N}c^2 +  E_\text{nuc} + n\frac{\partial E_\text{nuc}}{\partial n_\textrm{p}}, \\
&\mu_\textrm{e} = \hbar c(3\pi^2 n_\textrm{e})^{1/3}, \\
&\mu_\mu = \sqrt{(m_\mu c^2)^2 + \hbar^2 c^2 (3\pi^2 n_\mu)^{2/3}},
\end{align}
where we have assumed $E_\text{nuc} = E_\text{nuc}(n_\textrm{n},\ n_\textrm{p})$. Note that because $\partial \mu_\textrm{n}/\partial n_\textrm{p} \neq 0$ and $\partial \mu_\textrm{p}/\partial n_\textrm{n} \neq 0$, even if we neglect entrainment (i.e., terms containing $\alpha$), neutrons and protons are still coupled through the equation of state (see also \citealt{Prix:02}).

Although we use Newtonian equations to describe the stellar structure and oscillations, we write the mass density as $\rho=\varepsilon/c^2$ (and not $\rho=(n_\textrm{n} + n_\textrm{p})m_\textrm{N}$) in order to capture the composition gradients that arise from the nuclear interaction energy $E_{\rm nuc}$ and lepton fraction gradients. If we write the total mass density as the sum of each particle species $\rho = \rho_\textrm{n} + \rho_\textrm{p} + \rho_\textrm{e} + \rho_\mu$, then
\begin{align}
&\rho_\textrm{n} = n_\textrm{n} \left(m_\textrm{N} + \frac{E_\textrm{nuc}}{c^2} \right), \\
&\rho_\textrm{p} = n_\textrm{p} \left(m_\textrm{N} + \frac{E_\textrm{nuc}}{c^2} \right), \\
&\rho_\textrm{e} = \frac{T_\textrm{e}}{c^2},\\
&\rho_\mu= \frac{T_\mu}{c^2}.
\end{align}

The generalized pressure function $P$ for a two-fluid problem can be defined through the usual enthalpy density $w$ as
\begin{equation}
\varepsilon + P = w = \sum_{j=\textrm{npe}\mu} \mu_j n_j. 
\end{equation}
This gives the differential form
\begin{align}
\diff P &= \sum_{j=\textrm{npe}\mu} n_j \diff\mu_j - \alpha \diff \vel{v}_\textrm{r}^2.
\end{align}

It is convenient to define the specific chemical potential 
\begin{equation}
\diff \tilde{\mu}_j=\frac{\diff \mu_j}{m_j},
\end{equation}
where $m_j=\rho_j/n_j$. Note that $m_j$ is not the usual rest mass of particle $j$ (in particular, it is a function of density). Our definition of $\diff \tilde{\mu}_j$ is slightly different from that used in \cite{Andersson:01} and \cite{Prix:02} who take $\rho=(n_\textrm{n} + n_\textrm{p})m_\textrm{N}$ because they do not focus on g modes induced by composition gradients. Nonetheless, if we approximate $\diff \tilde{\mu}_\textrm{n}$ (which is the only specific chemical potential that explicitly enters our numerical calculations; see appendix \ref{subsec:oscEq}) as $\diff \mu_\textrm{n}/m_\textrm{N}$, it only changes our results at the few percent level. 

In our analytic work, it is also convenient to introduce a chemical potential $\tilde{\mu}_\textrm{c}$ corresponding to the normal fluid component of the fluid and defined such that
\begin{equation}
\rho_\textrm{c} \diff \tilde{\mu}_\textrm{c} =  \rho_\textrm{p} \diff \tilde{\mu}_\textrm{p} +\rho_\textrm{e} \diff\tilde{\mu}_\textrm{e} + \rho_\mu \diff\tilde{\mu}_\mu, 
\label{eq:tilde_mu_c_def}
\end{equation}
where $\rho_\textrm{c}=\rho_\textrm{p}+ \rho_\textrm{e} + \rho_\mu$. Note that $\tilde{\mu}_\textrm{c}$ is not itself an independent variable, but rather a function $\tilde{\mu}_\textrm{c}= \tilde{\mu}_\textrm{c}(\tilde{\mu}_\textrm{p}, \,\tilde{\mu}_\textrm{e}, \,\tilde{\mu}_\mu)$. Moreover, our calculation of the background model and the set of oscillation equations we solve numerically do not depend on $\tilde{\mu}_\textrm{c}$; we explicitly use $\tilde{\mu}_\textrm{c}$ only in Section \ref{subsec:oscEq} when manipulating the set of differential equations defining the linear perturbation operator $\mathcal{L}$ . 

Given the definitions above, we have
\begin{equation}
\diff P = \rho_\textrm{n} \diff \tilde{\mu}_\textrm{n} + \rho_\textrm{c} \diff \tilde{\mu}_\textrm{c} -\alpha \diff \vel{v}^2.
\label{eq:diffPdef}
\end{equation}
In hydrostatic and beta equilibrium, this implies
\begin{equation}
\frac{\diff \tilde{\mu}_\textrm{n}}{\diff r} + \frac{\diff \Phi}{\diff r} = 0,
\end{equation}
(up to small corrections due to leptonic contribution to the mass density). Note that this relation only holds in the static background and not in an oscillating fluid element.

Furthermore, if we define the deviation from beta equilibrium as 
\begin{equation}
\diff \beta = \diff \tilde{\mu}_\textrm{c} - \diff \tilde{\mu}_\textrm{n},
\end{equation}
then equation {\ref{eq:diffPdef}} implies \citep{Andersson:01}
\begin{align}
\label{eq:dmun_dP}
&\frac{1}{\rho} = \left(\frac{\partial \tilde{\mu}_\textrm{n}}{\partial P}\right)_\beta, \\
\label{eq:dmun_dbeta}
&\frac{\rho_\textrm{c}}{\rho} = -\left(\frac{\partial \tilde{\mu}_\textrm{n}}{\partial \beta }\right)_P, \\
\label{eq:drho_dbeta}
&\rho^2\frac{\partial}{\partial P}\left(\frac{\rho_\textrm{c}}{\rho}\right)_\beta = \left(\frac{\partial \rho}{\partial \beta}\right)_P.
\end{align}
These relations are used in Appendix \ref{subsec:oscEq} when we manipulate the oscillation equations in order to express them in a form convenient for proving the Hermiticity of $\mathcal{L}$.

\subsection{Entrainment function}\label{subsec:Entrainment}
The entrainment function $\alpha$ accounts for the `drag' between the superfluid neutrons and the protons when they are in relative motion (see equation \ref{eq:thermoDynamicID}). Many studies have discussed the entrainment effect in the context of oscillations of superfluid NSs (see, e.g., \citealt{Lindblom:94}, \citealt{Lee:95}, \citealt{Andersson:01}, \citealt{Prix:02}, \citealt{Kantor:14}, \citealt{Passamonti:16}, and \citealt{Dommes:16}). Most of the discussions originate from the study by \cite{Andreev:76}, who parametrize the entrainment effect in terms of the Landau effective masses of neutrons and protons, $m_\textrm{n}^\ast$ and $m_\textrm{p}^\ast$.  However, different authors adopt different notational conventions; the purpose of this appendix is to provide the connection between our notation and that of other studies.

Following \cite{Andersson:01} and \cite{Prix:02}, we parameterize $\alpha$ as
\begin{equation}
2\alpha=\frac{\left(m_\text{N} - m_\textrm{p}^\ast\right)\rho_\textrm{c}}{m_\text{N} + x_p(m_\text{N} - m_\textrm{p}^\ast)}
\end{equation}
and define the dimensionless entrainment functions
\begin{align}
&\epsilon_\textrm{n}=\frac{2\alpha}{\rho_\textrm{n}}, \\
&\epsilon_\textrm{c}=\frac{2\alpha}{\rho_\textrm{c}}=\frac{\rho_\textrm{n}}{\rho_\textrm{c}}\epsilon_\textrm{n}.
\end{align} 
Typical values of $m_\textrm{p}^{\ast}$ are in the range $0.3\le m_\textrm{p}^\ast/m_\textrm{N} \le 0.8$ \citep{Sjoberg:76, Chamel:08}; the smaller the $m_\textrm{p}^\ast$ is the greater $\alpha$ is and the stronger the entrainment effect is. 

\cite{Lindblom:94} and \cite{Lee:95} describe the entrainment effect through a mass density matrix $\rho_{ij}$ which relates the mass current and the macroscopically averaged velocities ($\vect{V_\textrm{c}}$, $\vect{V_\textrm{n}}$):
\begin{equation}
\left(\begin{array}{c}
\rho_\textrm{c}\vect{\vel{v}}_\textrm{c} \\ 
\rho_\textrm{n}\vect{\vel{v}}_\textrm{n}
\end{array}\right)=
\left(
\begin{array}{cc}
\rho_\textrm{cc} & \rho_\textrm{cn} \\
\rho_\textrm{nc} & \rho_\textrm{nn}
\end{array}
\right)
\left(\begin{array}{c}
\vect{V_\textrm{c}} \\ 
\vect{V_\textrm{n}}
\end{array}\right).
\end{equation}
Note that ($\vect{V_\textrm{c}}$, $\vect{V_\textrm{n}}$) are different from the microscopic velocities ($\vect{\vel{v}_\textrm{c}}$, $\vect{\vel{v}_\textrm{n}}$) we use here, which follow the definitions in \citet{Andersson:01} and \citeauthor{Prix:02} (2002; see the discussion in Appendix A2 of \citealt{Andersson:01}). The elements of $\rho_{ij}$ satisfy
\begin{align}
&\rho_\textrm{cc}+\rho_\textrm{cn}=\rho_\textrm{c}, \\
&\rho_\textrm{nc}+\rho_\textrm{nn}=\rho_\textrm{n}, \\
&\rho_\textrm{cn} = \rho_\textrm{nc}.
\end{align}
In terms of $m_\textrm{p}^\ast$ and $m_\textrm{n}^\ast$ \citep{Andreev:76}
\begin{align}
&\rho_\textrm{cc} = \rho_\textrm{c}\frac{m_\textrm{N}}{m_\textrm{p}^\ast}, \\ 
&\rho_\textrm{nn}=\rho_\textrm{n}\frac{m_\textrm{N}}{m_\textrm{n}^\ast},  \\
&\rho_\textrm{cn} = \rho_\textrm{c} \frac{m_\textrm{p}^\ast-m_\textrm{N}}{m_\textrm{p}^\ast} = \rho_\textrm{n} \frac{m_\textrm{n}^\ast-m_\textrm{N}}{m_\textrm{n}^\ast}.
\end{align}
Two useful relations connecting our notation to $\rho_{ij}$ are
\begin{align}
&2\alpha = -\frac{\rho_\textrm{c} \rho_\textrm{n}}{\det \rho}\rho_\textrm{cn}, \\
&1-\epsilon_\textrm{n}-\epsilon_\textrm{c}=\frac{\rho_\textrm{c} \rho_\textrm{n}}{\det \rho},
\end{align}
where $\det \rho = \rho_\textrm{cc}\rho_\textrm{nn} - \rho_\textrm{cn}^2$ is the determinant of $\rho_{ij}$.

\cite{Gusakov:14}, \cite{Dommes:16}, and \cite{Passamonti:16} take finite-temperature and general-relativistic effects into account. Nonetheless, their notations can be connected to ours in the appropriate zero-temperature, Newtonian limit. In this limit, the $y$ parameter used in \cite{Gusakov:14} and \cite{Dommes:16} is given by
\begin{equation}
y\simeq\frac{1}{x_\textrm{n}}\left(x_\textrm{p} - \epsilon_\textrm{n}\right),
\end{equation}
while the entrainment coefficient $\beta_\text{PAH}$ defined in \cite{Passamonti:16}  is given by  \begin{equation}
\beta_\text{PAH} \simeq 1 - \frac{2\alpha}{\rho_\textrm{n}} = 1-\epsilon_\textrm{n}
\end{equation}
(here we include a `PAH' subscript to distinguish it from the variable $\beta$ we use elsewhere and define as $\diff \beta=\diff \tilde{\mu}_\textrm{c}-\diff \tilde{\mu}_\textrm{n}$).

\section{superfluid oscillation equations and hermiticity  of the linear operator}\label{sec:oscEq}
In this Appendix we describe the superfluid oscillation equations in further detail. In Section \ref{subsec:oscEq}we present the form of the Newtonian oscillation equations that we use in our numerical calculations and in Section  \ref{subsec:boundCond} we describe the boundary conditions that we assume.  Our mode decomposition (eq. \ref{eq:mode_decomp}) relies on the linear operator $\mathcal{L}$ of the oscillation equations being Hermitian, which we prove in Appendix \ref{subsec:hermiticity}.

\subsection{Oscillation equations}\label{subsec:oscEq}
As we are considering a two-fluid problem, we need to consider the continuity and momentum conservation of both the charged flow and the neutron flow, which are given respectively by \citep{Prix:02}
\begin{align}
&\partial_t \rho_\textrm{c} + \vect{\nabla} \cdot \left(\rho_\textrm{c} \vect{\vel{v}_\textrm{c}}\right) = 0, \\
&\partial_t \rho_\textrm{n} + \vect{\nabla} \cdot \left(\rho_\textrm{n} \vect{\vel{v}_\textrm{n}}\right) = 0, \\
&\left(\partial_t + \vect{\vel{v}_\textrm{c}} \cdot \vect{\nabla}\right)\left(\vect{\vel{v}_\textrm{c}} - \epsilon_\textrm{c} \vect{\vel{v}_r}\right) -\epsilon_\textrm{c} \vel{v}_{r,i}\nabla \vel{v}_\textrm{c}^i= -\vect{\nabla}\left(\tilde{\mu}_\textrm{c}+\Phi\right), \\
&\left(\partial_t + \vect{\vel{v}_\textrm{n}} \cdot \vect{\nabla}\right)\left(\vect{\vel{v}_\textrm{n}} + \epsilon_\textrm{n} \vect{\vel{v}_r}\right) + \epsilon_\textrm{n} \vel{v}_{r,i}\nabla \vel{v}_\textrm{n}^i= -\vect{\nabla}\left(\tilde{\mu}_\textrm{n}+\Phi\right).
\end{align}
The set of equations is closed by the Poisson equation
\begin{equation}
\nabla^2 \Phi = 4\pi G(\rho_\textrm{c} + \rho_\textrm{n}).
\end{equation}

We use $\delta$ to denote Eulerian perturbations and assume all perturbed quantities have an  $e^{i\sigma t}$ time dependence. The Lagrangian displacements of the charged and neutron flows are thus given by
\begin{align}
&\partial_t \vect{\xi_\textrm{c}} = i \sigma \vect{\xi_\textrm{c}} = \delta \vect{\vel{v}_\textrm{c}}, \\
&\partial_t \vect{\xi_\textrm{n}} = i \sigma \vect{\xi_\textrm{n}} = \delta \vect{\vel{v}_\textrm{n}}. 
\end{align}
We further simplify the equations by assuming a spherical, hydrostatic background star and separating the variables into radial and angular functions using the standard spherical harmonic decomposition  (eqs. \ref{eq:SHdecompScalar} and \ref{eq:SHdecompVect}).  The linearized equations then reduce to a set of coupled ordinary differential equations in the radial direction
\begin{align}
&\frac{1}{r^2}\frac{\diff}{\diff r}\left(r^2\xi_\textrm{c}^r\right) + \frac{\diff \ln \rho_\textrm{c}}{\diff r} \xi_\textrm{c}^r - l\left(l+1\right)\frac{\xi_\textrm{c}^h}{r} + \frac{\delta \rho_\textrm{c}}{\rho_\textrm{c}} = 0,\\
&\frac{1}{r^2}\frac{\diff}{\diff r}\left(r^2\xi_\textrm{n}^r\right) + \frac{\diff \ln \rho_\textrm{n}}{\diff r} \xi_\textrm{n}^r - l\left(l+1\right)\frac{\xi_\textrm{n}^h}{r} + \frac{\delta \rho_\textrm{n}}{\rho_\textrm{n}} = 0,
\label{eq:nflow1}\\
&\sigma^2\left[\xi_\textrm{c}^r-\epsilon_\textrm{c}\left(\xi_\textrm{c}^r - \xi_\textrm{n}^r\right)\right] = \frac{\diff}{\diff r}\left(\delta \tilde{\mu}_\textrm{c} + \delta \Phi\right), \\
&\sigma^2\left[\xi_\textrm{n}^r+\epsilon_\textrm{n}\left(\xi_\textrm{c}^r - \xi_\textrm{n}^r\right)\right] = \frac{\diff}{\diff r}\left(\delta \tilde{\mu}_\textrm{n} + \delta \Phi\right), 
\label{eq:nflow2}\\
&\sigma^2\left[\xi_\textrm{c}^h-\epsilon_\textrm{c}\left(\xi_\textrm{c}^h - \xi_\textrm{n}^h\right)\right] = \frac{1}{r}\left(\delta \tilde{\mu}_\textrm{c} + \delta \Phi\right), \\
&\sigma^2\left[\xi_\textrm{n}^h+\epsilon_\textrm{n}\left(\xi_\textrm{c}^h - \xi_\textrm{n}^h\right)\right] = \frac{1}{r}\left(\delta \tilde{\mu}_\textrm{n} + \delta \Phi\right),
\label{eq:nflow3}\\
&\frac{1}{r^2}\frac{\diff}{\diff r}\left(r^2\frac{\diff \delta \Phi}{\diff r}\right) - \frac{l\left(l+1\right)}{r^2}\delta \Phi = 4\pi G\left(\delta \rho_\textrm{c}+\delta \rho_\textrm{n}\right).
\label{eq:dpoisson}
\end{align}
Since we have factored out the time-dependency by assuming perturbations vary as $e^{i \sigma t}$, we can write $\partial /\partial r$ as $\diff /\diff r$. 

For numerical reasons, it is convenient to define a `mass-averaged' flow 
\begin{equation}
\label{eq:xiplus}
\vect{\xi}_+  = \frac{1}{\rho} \left(\rho_\textrm{c} \vect{\xi}_\textrm{c} + \rho_\textrm{n}\vect{\xi}_\textrm{n}\right).
\end{equation}
The corresponding continuity and momentum conservation equations are then
\begin{align}
&\frac{\diff \xi_+^r}{\diff r} + \left(\frac{2}{r}+ \frac{\diff \ln \rho}{\diff r}\right)\xi_+^r - l(l+1)\frac{\xi_+^h}{r} + \frac{\delta \rho}{\rho}=0, 
\label{eq:avgFlow1}\\
&\sigma^2 \xi_{+}^r  = \frac{1}{\rho}\frac{\diff \delta P}{\diff r} + g\frac{\delta \rho}{\rho} +\frac{\diff \delta \Phi}{\diff r},
\label{eq:avgFlow2}\\
&\sigma^2 \xi_+^h=\frac{1}{r}\left(\frac{\delta P}{\rho} + \delta \Phi \right), \label{eq:avgFlow3}
\end{align}
where by equation (\ref{eq:diffPdef})
\begin{equation}
\delta P = \rho_\textrm{n} \delta \tilde{\mu}_\textrm{n} +  \rho_\textrm{c} \delta \tilde{\mu}_\textrm{c}.
\label{eq:deltaP}
\end{equation}
Summarizing, the set of oscillation equations we use to find numerical solutions are equations (\ref{eq:nflow1}, \ref{eq:nflow2}, \ref{eq:nflow3}, \ref{eq:dpoisson}, \ref{eq:avgFlow1}, \ref{eq:avgFlow2}, \ref{eq:avgFlow3}) and our independent variables are $(\xi_\textrm{n}^r, \xi_\textrm{n}^h, \xi_+^r, \xi_+^h, \delta P, \delta \tilde{\mu}_\textrm{n}, \delta \Phi)$. Following \cite{Kantor:14} and \cite{Passamonti:16}, we use ($P$, $\tilde{\mu}_\textrm{n}$, $x_{\mu \rm e}$) to parametrize the equation of state in the perturbed superfluid NS. For dependent variables $\delta \rho$ and $\delta \rho_\textrm{n}$ appearing in the equations, we project them onto the independent ones through the Jacobian
\begin{align}
\delta\rho&=\left(\frac{\partial \rho}{\partial P}\right)_{\tilde{\mu}_\textrm{n},x_{\mu\rm e}} \delta P + \left(\frac{\partial \rho}{\partial \tilde{\mu}_\textrm{n}}\right)_{P, x_{\mu\rm e}}\delta \tilde{\mu}_\textrm{n}+\left(\frac{\partial \rho}{\partial x_{\mu \rm e}}\right)_{P, \tilde{\mu}_\textrm{n}}\delta x_{\mu\rm e}, \nonumber \\
&= \left(\frac{\partial \rho}{\partial P}\right)_{\tilde{\mu}_\textrm{n},x_{\mu\rm e}}\delta P + \left(\frac{\partial \rho}{\partial \tilde{\mu}_\textrm{n}}\right)_{P, x_{\mu\rm e}}\delta \tilde{\mu}_\textrm{n}\nonumber \\
&\quad - \left(\frac{\partial \rho}{\partial x_{\mu \rm e}}\right)_{P, \tilde{\mu}_\textrm{n}}\frac{\diff x_{\mu \rm e}}{\diff r}\xi_\textrm{c}^r,
\end{align}
where in the second line we use the fact that the Lagrangian perturbation  $\Delta x_{\mu \rm e}$ vanishes because electrons and muons move at the same speed in the charged flow and therefore
\begin{equation}
\Delta x_{\mu \rm e}=\delta x_{\mu \rm e} + \frac{\diff x_{\mu \rm e}}{\diff r}\xi_\textrm{c}^r=0.
\end{equation}
We compute $\delta \rho_\textrm{n}$ through a similar expansion. Finally, the $\vect{\xi}_c$ terms are expressed in terms of ($\vect{\xi}_+$, $\vect{\xi}_\textrm{n}$) via equation (\ref{eq:xiplus}).

In addition to $\vect{\xi}_+$, in the main text we also introduce the displacement
\begin{equation}
\vect{\xi}_-  = (1-\epsilon_\textrm{n} - \epsilon_\textrm{c})(\vect{\xi}_\textrm{c} - \vect{\xi}_\textrm{n})=\frac{\rho_\textrm{c}\rho_\textrm{n}}{\det \rho}(\vect{\xi}_\textrm{c}-\vect{\xi}_\textrm{n}),
\end{equation}
which represents the difference between the normal fluid flow and the superfluid flow.  Although we do not use $\vect{\xi}_-$ when numerically solving the oscillation equations, it is useful for proving the Hermiticity of the linear perturbation operator $\mathcal{L}$ (Appendix \ref{subsec:hermiticity}).  Using ($\vect{\xi}_+$, $\vect{\xi}_-$) and defining $\delta \beta = \delta \tilde{\mu}_\textrm{c} - \delta \tilde{\mu}_\textrm{n}$, we can recast the oscillation equations as (see \citealt{Lindblom:94}, \citealt{Andersson:01}, and equations \ref{eq:dmun_dP}-\ref{eq:drho_dbeta})
\begin{align}
&\delta \rho + \vect{\nabla} \cdot (\rho \vect{\xi}_+)=0, \label{eq:xi+cont_vect_form}\\
&\left(\frac{\partial \rho}{\partial \beta}\right)_P\left(\frac{\delta P}{\rho}\right)+\frac{\rho_\textrm{n}^2}{\rho}\frac{\partial }{\partial \beta}\left(\frac{\rho_\textrm{c}}{\rho_\textrm{n}}\right)_P\delta \beta \nonumber\\ 
&\quad + \frac{1}{\rho}\left(\frac{\partial \rho}{\partial \beta}\right)_P \vect{\xi}_+\cdot\vect{\nabla}P + \vect{\nabla} \cdot\left(\tilde{\rho}\vect{\xi}_-\right)=0,\label{eq:LMterm} \\
&\frac{\partial^2 \vect{\xi}_+}{\partial t^2} =- \vect{\nabla}\left(\frac{\delta P}{\rho} + \delta \Phi \right) +\frac{\vect{\nabla}P}{\rho^2}\left(\frac{\partial \rho}{\partial \beta}\right)_P\delta \beta , 
\label{eq:xi+tt}\\
&\frac{\partial^2 \vect{\xi}_-}{\partial t^2} =- \vect{\nabla}\delta \beta,
\label{eq:xi-tt}
\end{align}
where 
\begin{equation}
\tilde{\rho} = \frac{\det \rho}{\rho} = \frac{\rho_\textrm{c} \rho_\textrm{n}}{(1-\epsilon_\textrm{n} - \epsilon_\textrm{c})\rho}.
\end{equation} 
Writing the oscillation equations in this form simplifies the proof of the Hermiticity of $\mathcal{L}$ given  in Appendix \ref{subsec:hermiticity}. Note that here we choose ($P$, $\beta$) to be the independent variables in our parameterization of the equation of state, and we use the relation 
\begin{equation}
\delta \rho = \left(\frac{\partial \rho}{\partial P}\right)_\beta\delta P + \left(\frac{\partial \rho}{\partial \beta}\right)_P\delta \beta.
\label{eq:drho_pbeta}
\end{equation}
As we discuss in Appendix \ref{subsec:basicQuan}, there are three independent variables when we parameterize an equation of state including muons. Indeed, $\delta\beta$ is a function of two independent variables since $\delta \beta=\delta \tilde{\mu}_\textrm{c}-\delta \tilde{\mu}_\textrm{n}$ and $\tilde{\mu}_\textrm{c}$ is a function of two independent variables [cf. equation \ref{eq:tilde_mu_c_def}; note that charge neutrality decreases the number of degrees of freedom by one]. 

\subsection{Boundary conditions}\label{subsec:boundCond}
The oscillation equations can be solved numerically when boundary conditions are specified. Here we focus on the set of equations described by the averaged flow ($\vect{\xi}_+$) and the superfluid neutron flow ($\vect{\xi}_\textrm{n}$), as they form the set of equations we solve numerically in practice. Other combination can be derived accordingly.

At the center ($r=0$) we apply the usual regularity condition 
\begin{align}
&\xi^r_+ = \xi^r_{0+} \frac{l}{\sigma^2}r^{l-1}, \\
&\xi^h_+=\xi^h_{0+} \frac{1}{\sigma^2}r^{l-1}, \\
&\xi^r_\textrm{n} = \xi^r_{0\rm n} \frac{l}{\sigma^2}r^{l-1}, \\
&\xi^h_\textrm{n} = \xi^h_{0\rm n} \frac{1}{\sigma^2}r^{l-1}, \\
&\delta P=\delta P_0 r^l, \\
&\delta \tilde{\mu}_\textrm{n}= \delta \tilde{\mu}_{n0} r^l, \\
&\delta \Phi= \delta \Phi_0 r^l, 
\end{align}
where
\begin{align}
&\xi^r_{0+}=\xi^h_{0+}, \\
&\xi^r_\textrm{0n}=\xi^h_\textrm{0n}, \\
&\xi^r_{0+} = \frac{\delta P_0}{\rho}  + \delta \Phi_0, \\
&\left[\left(1-\epsilon_\textrm{n} \frac{\rho}{\rho-\rho_\textrm{n}}\right)\xi^r_{0\rm n} + \epsilon_\textrm{n}\frac{\rho}{\rho-\rho_\textrm{n}}\xi^r_{0+}\right] = \delta \tilde{\mu}_\textrm{n} + \delta \Phi_0,
\end{align}
and all the background quantities are evaluated at $r=0$.

At the core -crust interface ($r=R_{\rm cc}$), we assume that the fluid becomes a normal fluid whose oscillation equations are identical to those of the averaged flow [equations \ref{eq:avgFlow1} - \ref{eq:avgFlow3}], but setting $\xi_+\to\xi_\text{NF}$, where $\xi_\text{NF}$ denotes the Lagrangian perturbation of normal fluid in the crust.  Continuity across the interface (from $R_{\rm cc}^-$ to $R_{\rm cc}^+$) then requires
\begin{align}
&\xi_+^r(R_\text{cc}^-)=\xi_\textrm{n}^r(R_\text{cc}^-)=\xi_\text{NF}^r(R_\text{cc}^+), \\
&\xi_+^h(R_\text{cc}^-)=\xi_\textrm{n}^h(R_\text{cc}^-)=\xi_\text{NF}^h(R_\text{cc}^+), \\
&\delta P(R_\text{cc}^-)=\delta P(R_\text{cc}^+), \\
&\delta \Phi(R_\text{cc}^-)=\delta \Phi(R_\text{cc}^+), \\
&\frac{\diff}{\diff r}\delta \Phi(R_\text{cc}^-) =  \frac{\diff}{\diff r} \delta \Phi(R_\text{cc}^+).
\end{align}

Finally at the surface ($r=R$), we require the Lagrangian perturbation of the pressure $\Delta P$ to vanish and the gravitational potential to be continuous, which gives (see, e.g., \citealt{Prix:02})
\begin{align}
&\Delta P=\delta P - \rho g \xi_\text{NF}^r=0, \\
&\frac{\diff \delta \Phi}{\diff r}+\frac{l+1}{r}\delta \Phi + 4 \pi G \rho \xi^r_\text{NF}=0.
\end{align}

\subsection{Hermiticity of linear perturbation operator}
\label{subsec:hermiticity}
In this section we prove that the linear perturbation operator $\mathcal{L}$ is Hermitian (see also \citealt{Lindblom:94, Andersson:04}). Let ($\vect{\xi}_+$, $\vect{\xi}_-$) and ($\vect{\xi}'_+$, $\vect{\xi}'_-$) denote two independent perturbations. We want to show that 
\begin{equation}
\left\langle 
\begin{bmatrix}
 \vect{\xi}_{+} \\ \vect{\xi}_{-}
\end{bmatrix},
\mathcal{L}
\begin{bmatrix}
\vect{\xi}'_{+} \\ \vect{\xi}'_{-} 
\end{bmatrix}\right\rangle = \left\langle 
\mathcal{L}
\begin{bmatrix}
 \vect{\xi}_{+} \\ \vect{\xi}_{-}
\end{bmatrix},
\begin{bmatrix}
\vect{\xi}'_{+} \\ \vect{\xi}'_{-} 
\end{bmatrix}\right\rangle 
\end{equation}
i.e., by equation (\ref{eq:innerProduct}),
\begin{align}
&\int \diff^3x\left[\rho\vect{\xi}^\ast_+ \cdot \mathcal{L} \left(\vect{\xi}'_+\right) + \tilde{\rho}\vect{\xi}^\ast_- \cdot \mathcal{L}\left(\vect{\xi}'_-\right)\right]
\nonumber \\
&=
\int \diff^3x\left[\rho \left\{\mathcal{L} \left(\vect{\xi}_+\right)\right\}^\ast \cdot\vect{\xi}'_+ + \tilde{\rho} \left\{\mathcal{L} \left(\vect{\xi}_-\right)\right\}^\ast \cdot \vect{\xi}'_-\right]
\end{align}
where $\mathcal{L}(\vect{\xi}_+)$ and $\mathcal{L}(\vect{\xi}_-)$ are given by the right hand sides of equations (\ref{eq:xi+tt}) and (\ref{eq:xi-tt}), respectively. Using equations (\ref{eq:xi+cont_vect_form})-(\ref{eq:xi-tt}) and defining
 \begin{equation}
 \delta W = \frac{\delta P}{\rho} + \delta \Phi,
 \end{equation}
 we have
 \begin{align}
&\int \diff^3x \rho \vect{\xi}^\ast_+ \cdot \mathcal{L} \left(\vect{\xi}'_+\right) \nonumber \\
&=\int \diff^3 x \rho \vect{\xi}^\ast_+\cdot\left[-\vect{\nabla} \delta W' + \frac{\vect{\nabla}P}{\rho^2}\left(\frac{\partial \rho}{\partial \beta}\right)_P\delta \beta' \right] \nonumber\\
&=\int \diff^3x \left[\vect{\nabla} \cdot (\rho \vect{\xi}^\ast_+)\delta W'  
+\frac{\delta \beta'}{\rho}\left(\frac{\partial \rho}{\partial \beta}\right)_P \vect{\xi}^\ast_+ \cdot \vect{\nabla}P\right]\nonumber \\
&= \int \diff^3 x\left[- \delta \rho^
\ast \delta W' + \frac{\delta \beta'}{\rho}\left(\frac{\partial \rho}{\partial \beta}\right)_P \vect{\xi}^\ast_+\cdot \vect{\nabla}P\right], 
\end{align}
and 
\begin{align}
&\int \diff^3x \tilde{\rho}\vect{\xi}^\ast_-\cdot\mathcal{L}\left( \vect{\xi}'_-\right)
= -\int \diff^3 x \tilde{\rho} \vect{\xi}^\ast_-\cdot \vect{\nabla}\delta \beta' \nonumber \\
&=\int \diff^3x \vect{\nabla} \cdot (\tilde{\rho} \vect{\xi}^\ast_-)\delta \beta' \nonumber \\
&=-\int \diff^3x \bigg[\left(\frac{\partial \rho}{\partial \beta} \right)_P\left(\frac{\delta P^\ast}{\rho}\right) + \frac{\rho_\textrm{n}^2}{\rho}\frac{\partial}{\partial \beta}\left(\frac{\rho_\textrm{c}}{\rho_\textrm{n}}\right)_P\delta \beta^\ast  \nonumber \\
&\quad  +\frac{1}{\rho} \left(\frac{\partial \rho}{\partial \beta}\right)_P\vect{\xi}^\ast_+ \cdot \vect{\nabla} P\bigg]\delta \beta',
\end{align}
where we have integrated by parts (the surface terms can be shown to vanish by the continuity relation at the core-crust interface and the assumption of vanishing surface density).
Adding the two equations together and using equation (\ref{eq:drho_pbeta}) we find
\begin{align}
&\int \diff^3x\left[\rho\vect{\xi}^\ast_+ \cdot \mathcal{L} \left(\vect{\xi}'_+\right) + \tilde{\rho}\vect{\xi}^\ast_- \cdot \mathcal{L}\left(\vect{\xi}'_-\right)\right] \nonumber \\
&=-\int \diff^3x\left[\left(\frac{\partial \rho}{\partial \beta}\right)_P\delta \beta^\ast + \left(\frac{\partial \rho}{\partial P}\right)_\beta \delta P^\ast\right]\delta W' \nonumber \\
&\quad -\int d^3x \left[\left(\frac{\partial \rho}{\partial \beta}\right)_P\left(\frac{\delta P^\ast}{\rho}\right) +\frac{\rho_\textrm{n}^2}{\rho}\frac{\partial }{\partial \beta}\left(\frac{\rho_\textrm{c}}{\rho_\textrm{n}}\right)_\beta \delta \beta^\ast \right] \delta \beta'  \nonumber \\
&=-\int d^3x\Bigg\{\left(\frac{\partial \rho}{\partial \beta}\right)_P\left[\delta \beta^\ast \delta W'+\delta \beta' \delta W^\ast\right] 
+ \rho \left(\frac{\partial \rho}{\partial P}\right)_\beta \delta W^\ast \delta W' \nonumber \\ 
&\quad + \frac{\rho_\textrm{n}^2}{\rho}\frac{\partial }{\partial \beta}\left(\frac{\rho_\textrm{c}}{\rho_\textrm{n}}\right)_P\delta \beta^\ast \delta \beta' -\rho\left(\frac{\partial \rho}{\partial P}\right)_\beta \delta \Phi^\ast \delta \Phi'\Bigg\} \nonumber \\ 
&\quad +\int \diff^3x \delta \Phi^\ast \left[\left(\frac{\partial \rho}{\partial P}\right)_\beta\delta P'+\left(\frac{\partial \rho}{\partial \beta }\right)_P\delta \beta' \right].
\end{align}
The last term can be rewritten using equations (\ref{eq:drho_pbeta}) and (\ref{eq:dpoisson}), which give
\begin{align}
&\left(\frac{\partial \rho}{\partial P}\right)_\beta\delta P'+\left(\frac{\partial \rho}{\partial \beta }_P\right)\delta \beta'  =\delta \rho'= \frac{\nabla^2\delta \Phi'}{4\pi G},
\end{align}
and noting that
\begin{equation}
\int \diff^3x \delta \Phi^\ast \nabla^2 \delta \Phi' = -\int \diff^3x \vect{\nabla}\delta \Phi^\ast \cdot \vect{\nabla}\delta \Phi',
\end{equation}
where the surface term vanishes. We thus prove that all the terms are symmetric under the exchange of $(\delta W^\ast, \delta \beta^\ast, \delta \Phi^\ast)$ and $(\delta W', \delta \beta', \delta \Phi')$, demonstrating that $\mathcal{L}$ is an Hermitian operator. 

\section{Numerical Accuracy of Tidal Coupling Coefficient Calculation}\label{sec:numErr}
\begin{figure}
\hspace*{-0.12cm}  
\includegraphics[angle=0,scale=.4]{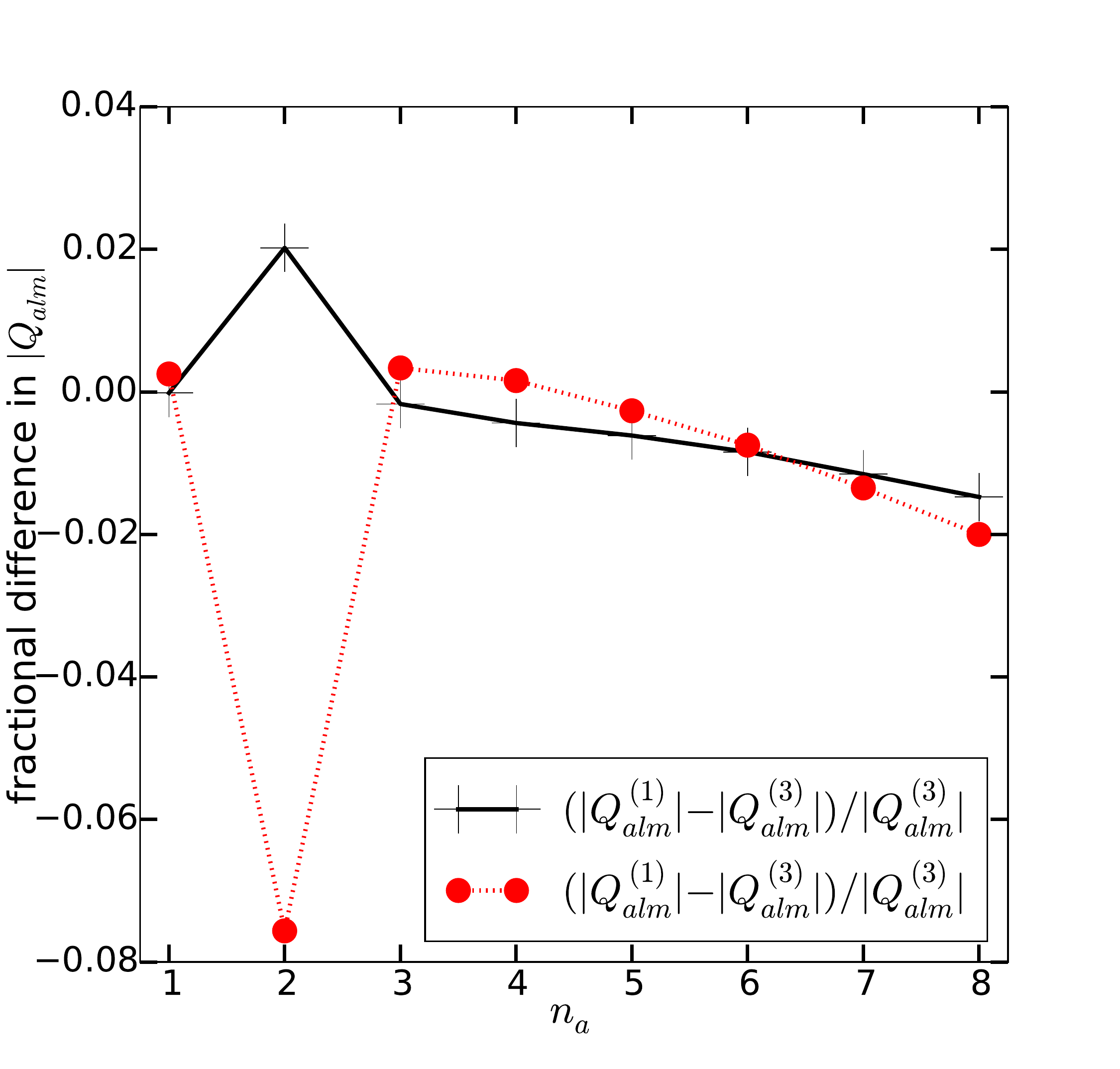}
\caption{Fractional difference between the three different methods of calculating $|Q_{alm}|$ (see text for details). The black solid lines are $|Q_{alm}^{(1)}-Q_{alm}^{(3)}|/|Q_{alm}^{(3)}|$ and the red dotted lines are $|Q_{alm}^{(2)}-Q_{alm}^{(3)}|/|Q_{alm}^{(3)}|$. }
\label{fig:numErr}
\end{figure}

The oscillatory nature of the g modes makes the calculation of the tidal coupling coefficient $Q_{alm}$ subject to numerical error \citep{Reisenegger:94b, Reisenegger:94,Weinberg:12}. We validated the accuracy of our calculations by evaluating $Q_{alm}$ in three different ways as given by equations  (\ref{eq:Qform_orig}) and (\ref{eq:coupCoefForms}):
\begin{align}
&Q_{alm}^{(1)}=\frac{1}{MR^l}\int \diff r l\rho r^{l+1}\left[\xi_{a+}^r+(l+1)\xi_{a+}^h\right],\\
&Q_{alm}^{(2)}=\frac{1}{MR^l}\int \diff r r^{l+2} \delta \rho_{a}, \\
&Q_{alm}^{(3)}=-\frac{2l+1}{4\pi} \frac{\delta \Phi_{a}(R)}{GM/R}. 
\end{align}
Although in the main text we give values based on $Q_{alm}^{(3)}$, we find that all three methods agree very well.  For example, in Fig. \ref{fig:numErr} we show the fractional differences $|Q_{alm}^{(1)}-Q_{alm}^{(3)}|/|Q_{alm}^{(3)}|$ and $|Q_{alm}^{(2)}-Q_{alm}^{(3)}|/|Q_{alm}^{(3)}|$ for the superfluid NS model with $(M/M_\odot,\ m_\textrm{p}^{\ast}/m_\text{N})=(1.4,\ 0.8)$. The differences are at the $\sim1$ per cent level.

\bsp	
\label{lastpage}

\end{document}